\documentstyle[aps,prl,preprint,epsfig]{revtex}
\begin{document}
\title{Nuclear binding energies: Global collective structure and 
local shell-model correlations}
\author{R.~Fossion$^a$, 
C.~De~Coster$^{a,}$\footnote{Postdoctoral fellow of the Fund for 
Scientific Research-Flanders (Belgium).},  
J.E.~Garc\'{\i}a-Ramos$^{a,b,}$\footnote{Visiting postdoctoral 
fellow of the Fund for Scientific Research-Flanders (Belgium).}, 
T.~Werner$^{a,c}$, and K.~Heyde$^{a,d}$}
\address{$^{a}$ Department of Subatomic and Radiation Physics,
Proeftuinstraat, 86 B-9000 Gent, Belgium}
\address{$^{b}$ Dpto.~de F\'{\i}sica Aplicada e Ingenier\'{\i}a
El\'ectrica, EPS La R\'abida, Universidad de Huelva, 21819 Palos de la
Frontera, Spain}
\address{$^{c}$ Institute for Nuclear 
Physics, University of Warsaw, Poland}
\address{$^{d}$ EP-ISOLDE, CERN, CH-1211, Geneva, Switzerland}
\title{Nuclear binding energies: Global collective structure and 
local shell-model correlations}

\maketitle

\begin{abstract}Nuclear binding energies and two-neutron separation 
energies are 
analysed starting from  the liquid-drop model and the nuclear shell
model in order to describe the global trends of the above observables.
We subsequently concentrate on the Interacting Boson Model (IBM) and 
discuss
a new method in order to provide a consistent description of both,
ground-state and excited-state properties. We address the artefacts
that appear when crossing mid-shell using the IBM formulation and perform
detailed numerical calculations for nuclei situated in the $50-82$ shell.
We also concentrate on local deviations from the above global trends in
binding energy and two-neutron separation energies that appear in the
neutron-deficient Pb region. We address possible effects on 
the binding energy, caused by mixing of low-lying $0^{+}$ intruder
states into the ground state, using configuration mixing in the IBM 
framework. We also
study ground-state properties using a macroscopic-microscopic model.
Detailed comparisons with recent experimental data in the Pb region are
amply discussed.
\end{abstract}

\vspace{2cm}


\noindent
{\small\bf Keywords: Binding energies, two-neutron separation 
energies,\\\phantom{Keywords:} liquid-drop model, 
shell model, interacting boson model,
\\\phantom{Keywords:} intruder states, macroscopic-microscopic model.}
\newpage

\section{Introduction}
\label{sec-intro}
In the study of nuclear structure properties, nuclear masses or
binding energies ($BE$) and, more
in particular, two-neutron separation energies ($S_{2n}$), are 
interesting probes to find out about specific nuclear structure 
correlations 
that are present in the nuclear ground state. 
These correlations, to a large extent, express the global
behavior that is most easily seen in a global way and as such,
the liquid drop model (LDM) serves 
as a first guide to match to the observed
trends concerning nuclear ground-state masses \cite{Ring80,Nils95}.
There have been extensive global mass studies carried out which aim,
in particular, at reproducing the overall trends: from pure liquid-drop
model studies (LDM) \cite{Waps58,Waps71,Myer67,Myer69}, 
over macroscopic-microscopic methods 
\cite{Moll81,Beng84,Moll81a,Moll88,Moll95,Moll97}, over the semi-classical
ETFSI (Extended Thomas-Fermi plus Strutinsky integral) \cite{Abou95},
towards, more
recently, relativistic mean-field \cite{Lala99}, and Skyrme
Hartree-Fock studies \cite{Gori01}.

It is our aim to concentrate on local correlations that are
rather small on the absolute energy scale used to describe binding 
energies (or two-neutron separation energies) but nevertheless 
point out to a number of interesting extra nuclear structure effects. 
These can come from various origins such as (i) the presence
of closed-shell discontinuities, (ii) the appearance of local zones of
deformation, and (iii) configuration mixing or shape mixing that shows
up in the nuclear ground state itself. Except
for the closed-shell discontinuities, the other effects give rise to
small energy changes, about $200-300$~keV or less, that were not
observed in experiments until recently.  
However, in the last few years a dramatic 
increase in the experimental sensitivity, using trap devices 
or specific mass-measurement set-ups (ISOLTRAP, MISTRAL, $\ldots$)
\cite{Boll96,Raim97}, has shifted the level of
accuracy down to a few tens of keV (typically $30-40$~keV for nuclei
in the Pb region) such that mass measurements are now of the level
to indicate local correlation energies that allow to test nuclear
models (shell-model studies) \cite{Schw98,Kohl99,Schw01}. 
Therefore, interest has  
been growing considerably and we aim at discussing and analysing, 
from this point of view, recent mass measurements.

In the first part of the present paper, we discuss the collective (or
global) features of the nuclear binding energy (or the $S_{2n}$
observable) using a simple liquid-drop model 
(section \ref{sec-ldm-be}) and
some general properties of the shell model (section
\ref{sec-shell-be}). In section \ref{sec-ibm-be} we concentrate on a
description of binding energies within the framework of the
Interacting Boson Model (IBM), where, besides a study of the global
aspects of $S_{2n}$, the specific nuclear structure correlations
(local part) are studied in, and close to, the symmetry
limits $U(5)$,
$SU(3)$, and $O(6)$. Special attention is given to obtain a consistent
description of $BE$ ($S_{2n}$) values when crossing the mid-shell
point. Applications for the $50-82$ shell are presented in some
detail. In the second part of the paper (section \ref{sec-intru}) we
concentrate on local modifications of the otherwise smooth $BE$
($S_{2n}$) behavior that come from the presence of low-lying intruder
states at and near closed shells. We discuss the effect in both, an
approximate IBM configuration mixing approach, as well as by studying,
in some detail, the macroscopic-microscopic model giving rise to
the potential energy surfaces (PES). We
apply and discuss both calculations for nuclei in the
neutron-deficient Pb region. Finally, in section \ref{sec-conclu} we
present a number of conclusions.   

\section{Liquid drop model behavior}
\label{sec-ldm-be}
Nuclear masses and the derived quantity, the two-neutron separation
energy, $S_{2n}$, form important indicators that may show the presence
of extra correlations on top of a smooth liquid-drop behavior. 
$S_{2n}$ is defined as
\begin{equation}
\label{s2n_be}
S_{2n}(A,Z)= BE(A,Z) - BE(A-2,Z), 
\end{equation}
where $BE(A,Z)$ is the binding energy defined as positive, 
({\it i.e.}~it is the positive of the energy of the ground state 
of the atomic nucleus) for a nucleus with $A$ nucleons and $Z$ protons.

The simplest LDM \cite{Waps58,Waps71} will be used as a reference 
point throughout this 
paper because it allows
the overall description of the {\em BE} along the whole table of 
masses, or for long series of isotopes. More sophisticated
macroscopic-microscopic finite-range droplet models, using a
folded-Yukawa single-particle potential have been developed
\cite{Moll81,Beng84} and extensive mass tables published
\cite{Moll81a,Moll88,Moll95,Moll97}. Here, the global trends, even passing
through regions of deformed nuclei, have been well
described. Deviations that result, usually point towards new local
nuclear structure effects. The present computing possibilities have
resulted in modern state-of-the-art mass tables spanning
semi-classical EFTSI methods \cite{Abou95}, over self-consistent
relativistic mean-field models \cite{Lala99}, towards a recent Skyrme
Hartree-Fock study \cite{Gori01}. Those mass calculations serve as
benchmarks for the whole nuclear mass region but local nuclear
structure correlations, as well as the exact location of the onset of
deformation stay outside the ``philosophy'' and scope of these
important studies.

\subsection{The global trend of $S_{2n}$ along the valley of
stability} 
\label{sec-ldm-vall}
\begin{figure}
\begin{center}
\mbox{\epsfig{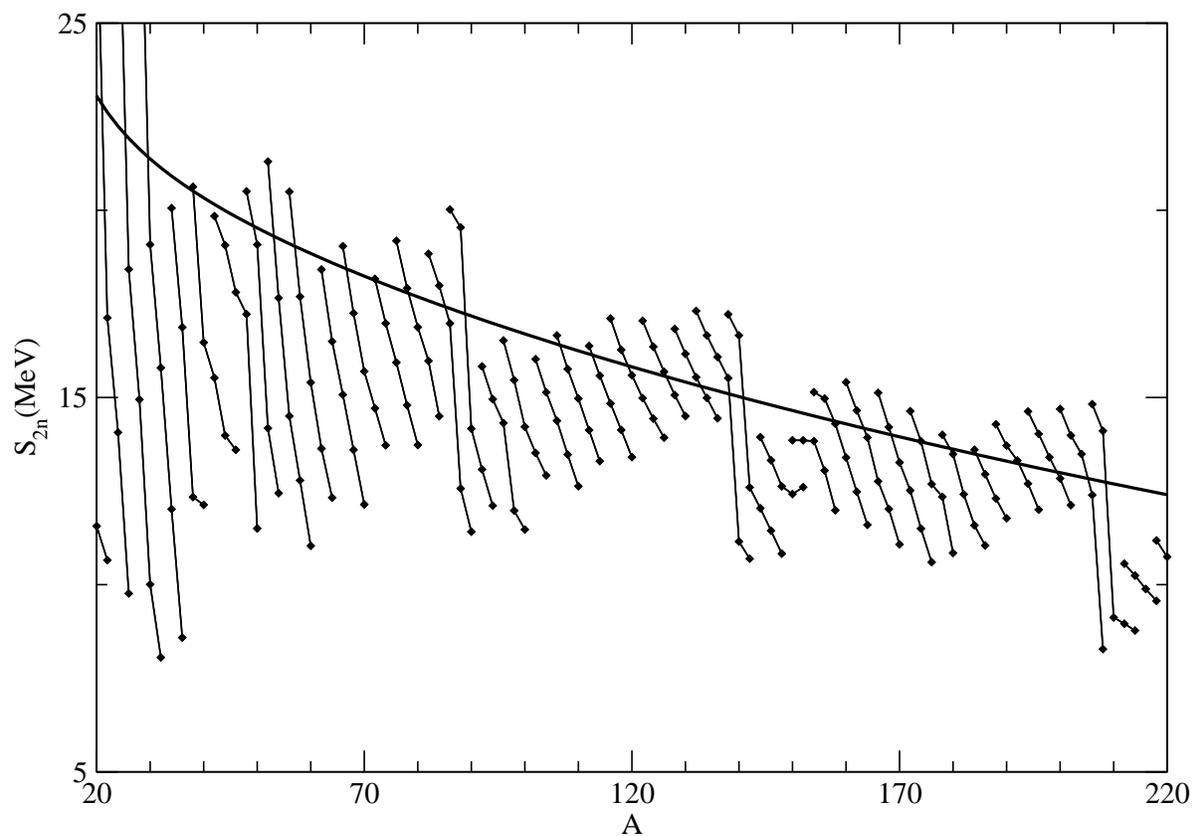}}
\end{center}
\caption{Comparison between the experimental $S_{2n}$ (diamonds) 
  values and the LDM prediction (thick full line) along the valley of  
  stability. The experimental data correspond to even-even nuclei around
  the line of maximum stability. Points connected with lines
  correspond to nuclei with equal $Z$. }
\label{fig-s2n-ldm-exp}
\end{figure}

To see how the values of $S_{2n}$ evolve globally, through the
complete mass chart, we start from can the semi-empirical mass 
formula \cite{Weiz35,Beth36}, 
\begin{equation} 
\label{ldm} 
BE(A,Z) = a_V A - a_S A^{\frac{2}{3}} - 
a_C Z (Z-1) A^{-\frac{1}{3}} - a_A (A-2Z)^2 A^{-1}.
\end{equation}
Even though there exist much refined macroscopic models (see
references given before), the present simple liquid-drop model
(\ref{ldm}) is a suitable starting point for our purpose in analysing
the physics behind the $S_{2n}$ systematics.
The $S_{2n}$ value can be written as,
\begin{equation}
S_{2n}\approx 
2(a_{V}-a_{A}) - \frac{4}{3} a_{S}A^{-\frac{1}{3}} + \frac{2}{3}
a_{C}Z(Z-1)A^{-\frac{4}{3}}+ 8a_{A}\frac{Z^2}
{A(A-2)},
\end{equation}
where the surface and Coulomb terms are only approximated expressions.  
If one inserts the particular value of $Z$, $Z_0$, 
that maximises the binding 
energy for each given $A$ (this is the definition 
for the valley of stability),
\begin{equation} 
Z_0 = \frac{A/2}{1+0.0077 A^{2/3}},
\end{equation} 
we obtain, for large values of $A$, the result (see \cite{Kohl99}),
\begin{equation}
\label{s2n-vs}
S_{2n}= 2(a_{V}-a_{A}) -\frac{4}{3}a_{S}A^{-\frac{1}{3}}
+(8a_{A}+\frac{2}{3}a_{C}A^{\frac{2}{3}})\frac{1}{4 + 0.06
A^{\frac{2}{3}}}. \label{s2n_ldm}
\end{equation}
In the present form, we use the following values 
for the LDM parameters: $a_{V}$ =
$15.85$ MeV, $a_{C}$ = $0.71$ MeV, $a_{S}$ = $18.34$ MeV and 
$a_{A}$ = $23.22$ MeV \cite{Waps58,Waps71}.  
In Fig.~\ref{fig-s2n-ldm-exp}, 
we illustrate the behavior of $S_{2n}$ along
the valley of stability (\ref{s2n-vs}) for even-even nuclei, 
together with the experimental
data. The experimental data correspond to a range of $Z$ between
$Z_0+1$ and $Z_0-2$. 
It appears that the overall decrease and the specific mass 
dependence is well contained within the liquid-drop model. 
   
\subsection{The global trend of $S_{2n}$ through a chain of isotopes}
\label{sec-ldm-local}
We can also see how well the experimental two-neutron 
separation energy, through a chain of isotopes, is reproduced using
the LDM. 
From Fig.~\ref{fig-s2n-ldm-exp}, it is clear that, besides the sudden
variations near mass number $A=90$ (presence of shell closure at
$N=50$) and near mass number $A=140$ (presence of the shell closure 
at $N=82$), the specific mass dependence 
for series of isotopes comes closer to specific
sets of straight lines.

\begin{figure}
\begin{center}
\mbox{\epsfig{file=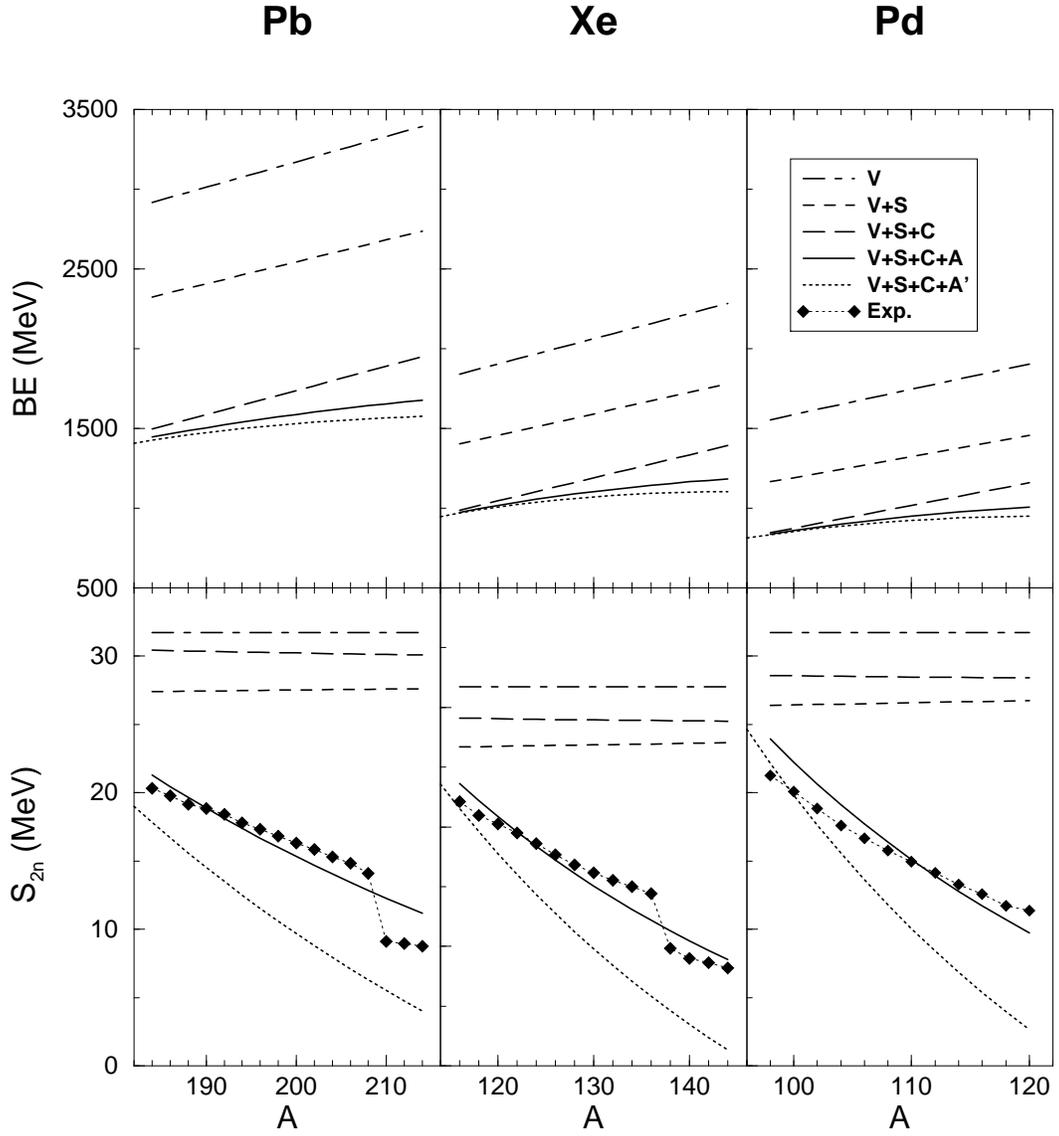,height=14.0cm,angle=-90}}
\end{center}
\caption{Contributions of the different terms of the mass formula to
  the $BE$ (top row) and $S_{2n}$ (bottom row) for Pb, Xe, and Pd. In
  the $S_{2n}$ panels are also shown the experimental data. Two
  different asymmetry terms are considered $a_A=23.22$ MeV ($A$), full
  line, and $a_A=30$ MeV ($A'$), dotted line.}
\label{fig-ldm-contrib}
\end{figure}

Next, we observe that the mass formula (\ref{ldm}) is able to
describe the observed almost linear behavior of $S_{2n}$ 
for series of isotopes. The more appropriate way of carrying out this 
analysis is to make an expansion of the different terms in (\ref{ldm})
around a particular value of $A$ (or $N$, because $Z$ is fixed), 
$A_0=Z+N_0$, and to keep the main orders. 
Therefore, we define $X=A-A_0$ and
$\varepsilon=X/A_0$. 
Let us start with the volume term,
\begin{equation} 
\label{ldm-v}
BE_V(A)-BE_V(A_0)=a_V X.
\end{equation} 
The surface term gives rise to,
\begin{equation} 
\label{ldm-s}
BE_S(A)-BE_S(A_0)\approx -a_S 
A_0^{\frac{2}{3}}\left(\frac{2}{3} \varepsilon 
-\frac{1}{9} \varepsilon^2 \right)
=
-a_S \frac{2}{3} {X\over A_0^{\frac{1}{3}}}+
a_S \frac{1}{9} {X^2\over A_0^{\frac{4}{3}}} ,
\end{equation} 
and the contribution of the Coulomb term is,
\begin{eqnarray} 
\label{ldm-c}
BE_C(A)-BE_C(A_0)&\approx& -a_C Z(Z-1) A_0^{\frac{1}{3}} 
\left(-\frac{1}{3} \varepsilon+\frac{2}{9}\varepsilon^2 \right)
\nonumber\\
&=& 
{a_C\over 3}Z(Z-1){X\over A_0^{\frac{4}{3}}}
-{2a_C\over 9}Z(Z-1){X^2\over A_0^{\frac{7}{3}}}.
\end{eqnarray} 
Finally, the asymmetry contribution is,
\begin{equation}
\label{ldm-a}
BE_A(A)-BE_A(A_0)\approx -a_A \left(A_0-{4Z^2\over A_0}
\right)\varepsilon-a_A{4Z^2\over A_0}\varepsilon=-
a_A\left(1-{4Z^2\over A_0^2}\right)X-
a_A{4Z^2\over A_0^3}X^2. 
\end{equation} 
First, it is clear that the coefficients of the linear part are (for 
$A_0\approx 100$ and $Z\approx 50$ and taking for $a_{V}$, $a_{C}$ 
$a_{S}$, and  $a_{A}$ the values given in previous section) 
about two orders of  magnitude larger than the coefficients of the 
quadratic contribution. With respect to the second order terms it is
interesting to see the value of each of them: the surface term
gives $0.0044$ MeV, the Coulomb term $-0.0083$ MeV, and the
asymmetry term $-0.23$ MeV. As a consequence, in this case, the leading
term is the asymmetry one and it is essentially the main source of
non-linearities in the $BE$ and therefore, the source of the slope of the
$S_{2n}$.  
In order to illustrate these results, we present in 
Fig.~\ref{fig-ldm-contrib} the
different contributions of the LDM (volume, volume plus surface, 
volume plus surface plus Coulomb and volume plus surface plus Coulomb
plus asymmetry term) to the $BE$ and $S_{2n}$ 
for different families of isotopes. It thus appears that only the 
asymmetry term induces the quadratic behavior in $BE$ and the linear
one in $S_{2n}$.

In eq.~(\ref{s2n-vs}) we make use of the classical values as discussed
in the early papers of Wapstra \cite{Waps58,Waps71}. This fit, of
course, was constrained to a relatively small set of experimental data
points. More recent fits, including (i) higher-order terms in the LDM,
{\it e.g.}~the surface symmetry correction, a finite-range surface
term, droplet correction, etc.~and (ii) the much extended range of
experimental data points, seem to give rise to an increased asymmetry
energy coefficient $a_A$ with values of the order of $\approx 30$ MeV
\cite{Krup00}. We show, in Fig.~\ref{fig-ldm-contrib}, 
the results obtained
using an increased $a_A$ coefficient, which gives rise to an increased
in the absolute value of the $S_{2n}$ slope.

In the present analysis, we have left out the effect of the pairing
energy, with a dependence $BE_{\mbox{pairing}}=-11.46 A^{-1/2}$
\cite{Waps58,Waps71}, because the net result is an overall shift in the binding
energy, but the relative variation in the $S_{2n}$ values, over a mass
span of $\Delta A=20$ units, is of the order of $\approx 100$ keV and,
as such, is not essential for the plot of
Fig.~\ref{fig-ldm-contrib}.    

Finally it should be stressed that, far from stability 
and for very neutron-rich nuclei, the asymmetry term can be at 
the origin of a decreasing trend in the $BE$ 
when $A$ further increases. This actually corresponds to the well
known drip-line phenomenon, but here within a pure LDM approach.

\section{Shell-model description of binding energies}
\label{sec-shell-be}
Within the liquid drop model (LDM) 
description of the binding energy  of
atomic nuclei, the volume term, the surface and Coulomb energy
contributions give rise to an essentially flat behavior in the
$S_{2n}$ values. It is the asymmetry term that accounts for an almost
linear drop in the quantity $S_{2n}$ within a given isotopic series
as a function of the nucleon (or neutron) number $A$ (or $N$). This
expresses  the progressively decreasing binding energy needed to 
remove a pair of nucleons out of a given nucleus.

The asymmetry term shows up already when treating the nucleus as a
Fermi gas of independent particles, as a consequence of the Pauli
principle. The linear drop in $S_{2n}$ also shows up in a more realistic
shell-model description of nucleons
moving in an average field, characterised 
by the single-particle spectrum
$\epsilon_{j}$, that are subsequently coupled to 
$J^{\pi}=0^{+}$ pairs because of the
major attractive binding-energy correlation on top of the monopole
binding-energy term. To fix the idea, one should start from a
doubly-closed shell nucleus as a reference nucleus in order to
describe binding energies (or separation energies) and then  treat the
interactions amongst nucleons filling a given single-$j$ shell. Talmi
has shown \cite{Shal63,Talm93} that, for a zero-range 
force ($\delta$-function
interaction) using a pair-coupled wave function that has seniority $v$
as a good quantum number, the contribution to the ground-state
configuration can be expressed as
\begin{equation} 
\label{be-delta}
BE(j,n) = \langle j^{n},v=0,J=0| \sum V(i,k) |j^{n},v=0,J=0\rangle,
\end{equation} 
or
\begin{equation} 
\label{be-delta2}
BE(j,n) = n \epsilon_{j} + \frac{n}{2} V_{0} ,
\end{equation} 
where $V_{0}=\langle j^{2},v=0,J=0|V|j^{2},v=0,J=0\rangle$ 
is the attractive $0^{+}$ two-body matrix element. 
This binding energy contribution is essentially
equal to the volume part of the liquid-drop model formulation (which
scales like $A$), and scales with the number of interacting valence
nucleons moving in the single-$j$ shell-model orbital and contributes 
with a constant value to the $S_{2n}$ two-neutron separation energy. 

\begin{figure}
\begin{center}
\mbox{\epsfig{file=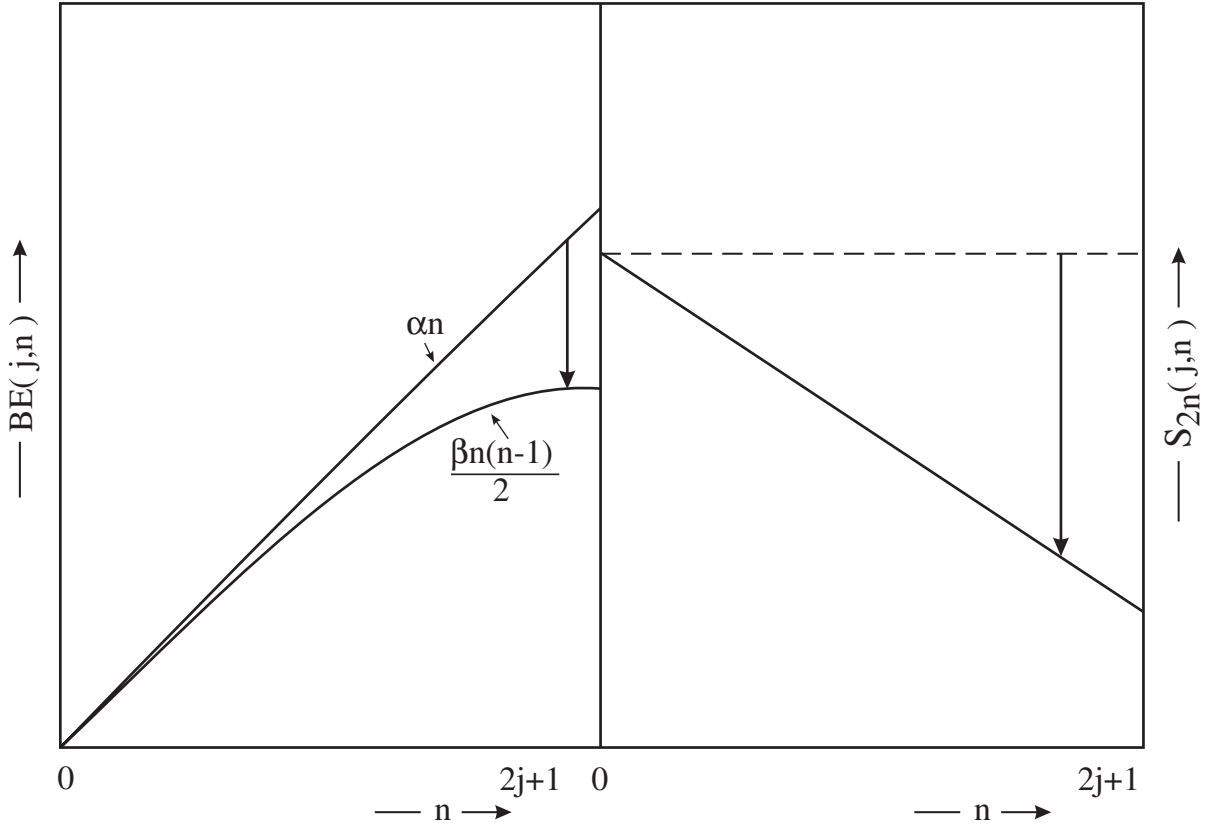,height=11.0cm,angle=0}}
\end{center}
\caption{Schematic representation of $BE$ (left) and $S_{2n}$ (right) 
  for a shell model Hamiltonian that preserves the seniority $v$ 
  as a good quantum number. The two different contributions to $BE$
  ($S_{2n})$ are plotted separately.}
\label{fig-be-sm}
\end{figure}

More general interactions (finite range forces, the standard pairing
force, \ldots) 
contribute with extra terms in the expression 
of the binding energy \cite{Shal63}.
Coupling of the ground-state seniority $v=0$ with higher-lying seniority
$v=2,4,\dots$ configurations also modifies this most simple 
expression given in
(\ref{be-delta2}). 
This then leads to a general diagonal 
energy that also contains a term quadratic in the number of 
nucleons $n$ (the specific coefficients
depend on the specific forces and coupling) and is given as
\cite{Talm71} 
\begin{equation} 
\label{be-sm}
BE(j,n)=C+\alpha n+\beta\frac{{n}{(n-1)}}{2}+[\frac{n}{2}]P, 
\end{equation} 
provided the seniority $v$ is a good quantum number. Here, $[n/2]$
stands for the largest integer not bigger than $n/2$. Further, $\alpha$
is in general large and attractive
($\alpha=\epsilon_{j}+\frac{V_{0}}{2}$ with $\epsilon\approx -8$ MeV and 
$\frac {V_{0}}{2} \approx -(j+ \frac {1}{2})$ $G$, with $G$ 
the pairing force strength), $\beta$ is much smaller
and repulsive (in agreement with the sign of the asymmetry term in the
liquid-drop energy expression) and $P$ describes the odd-even pairing
staggering to the binding-energy expression (see
Fig.~\ref{fig-be-sm}). 
Thus, it is the $\beta$
contribution that causes a linear drop in the $S_{2n}$ value as a function
of the nucleon number. This expression has been 
used to fit $S_{2n}$ values
in various mass regions \cite{Talm93,Talm71,Wood-p}. 
This shell-model behavior 
actually describes the long stretches of linear behavior in the $S_{2n}$
curve over a large region of the nuclear mass table, indicating that the
above simple structure contains the correct physics and saturation
properties of the nucleon-nucleon two-body interactions. From the above
discussion, it becomes clear that, in order to correctly reproduce the
experimental $S_{2n}$ behavior over a large series of isotopes, one
needs a good description of the single-particle energies $\epsilon_{j}$
and their variation over a given mass region. The monopole term
\cite{Dufo96,Dufl99} is essential  
in order to correctly reproduce saturation
in the nuclear binding starting from a pure 
shell-model approach, {\it i.e.}~if one starts out, 
{\it e.g.}~with the single-particle energies in the $sd$
shell-model space around $^{16}$O and considers the variation in these
single-particle energies through the monopole proton-neutron contribution
\begin{equation} 
\label{lev-corr}
\tilde{\epsilon}_{j_{\rho}}=\epsilon_{j_{\rho}}+
\sum v^{2}_{j_{\rho '}}\langle j_{\rho}
j_{\rho '}|V|j_{\rho}j_{\rho '}\rangle,
\end{equation} 
one should reproduce the observed relative energy spacing in the $sd$
shell when reaching the end of the shell near $^{40}$Ca.

We shall not start to discuss detailed shell-model calculations here but
we like to refer to the very good reproduction of the overall behavior
in the $S_{2n}$ value when crossing the full {\em sd} shell-model region 
\cite{Brow88}, except near $N=20$ and for the Ne, Na, and Mg nuclei
\cite{Warb90,Wood92}.

In the next section, we shall carry out a more detailed study of $S_{2n}$
properties with a shell-model space that is truncated to that part which
mainly determines the low-lying collective properties, 
{\it i.e.}~we will perform
the IBM symmetry truncation. We expect essentially to recover the
shell-model features as described here. The IBM will allow, however,
a more detailed study covering large sets of isotopes in the nuclear
mass table.

\section{IBM description of binding energies}
\label{sec-ibm-be}
The Interacting Boson Model \cite{Iach87}
takes advantage of the group theory 
for describing low-lying states of even-even nuclei. 
Such states present a clear {\it quadrupole} collectivity. 
The building blocks of the model are bosons with angular momentum
$L=0$ ($s$ bosons) and angular momentum $L=2$ ($d$ bosons). The number of
interacting bosons that are present in the system corresponds to half
the number of valence nucleons, $N=n/2$, and they interact through a
Hamiltonian containing, in the simplest case, up to two-body
interactions, being number conserving and rotationally invariant. The
original version of the model is called IBM-1 and in this approach no 
difference is considered between protons and neutrons
\cite{Iach87,Fran94}. In this
section we use the IBM-1 version of the model.

In the last few decades the IBM has provided a satisfactory
description of spectra and transitions rates of medium-mass and heavy
nuclei \cite{Cast88}. However in most of the cases the binding 
energy has not been
considered in the analysis. In a recent paper \cite{Garc01a}
it was pointed out that it
is extremely important to include the $BE$, or equivalently the
$S_{2n}$ values, in an IBM study because 
its value is very sensitive to the 
Hamiltonian that it is used. Therefore, it is very useful for choosing
the most appropriate Hamiltonian in the description of a given
nucleus. As was shown by Garc\'{\i}a-Ramos {\it et al.} in  
Ref.~\cite{Garc01a} and will be recapitulated here, 
in order to study the binding-energy properties,
it is necessary to analyse a complete chain
of isotopes and not a single nucleus, which makes the study more
complicated.

At this point it is convenient to write the definition of
$S_{2n}$. In case we use nucleon particles outside of a closed shell,
we define (remind, $N$ denotes the number of nucleon pairs outside of
closed shells, not to be confused with the neutron number):
\begin{equation} 
\label{sn2-def1}
S_{2n}=BE(N)-BE(N-1).
\end{equation} 
When using a description in terms of holes inside a closed shell, the
definition of $S_{2n}$ becomes,
\begin{equation}
\label{s2n-def2}
S_{2n}=BE(N)-BE(N+1).
\end{equation}
Later (section \ref{sec-mid}), we shall present
a prescription that contains only a single definition.
 
For later use (section \ref{sec-real}) 
we present here a very compact 
IBM Hamiltonian that will be used throughout this section. 
This Hamiltonian is not the most general one
but is ideal for the purpose of studying binding energies and allows 
to describe many
realistic situations \cite{Chou97}. It can be written as follows,
\begin{equation} 
\label{ham1}
\hat H=\epsilon_d \hat n_d-\kappa\hat Q\cdot\hat Q+
\kappa' \hat L\cdot\hat L, 
\end{equation}
where $\hat n_d$ is the $d$ boson number operator and
\begin{eqnarray}
\label{L}
\hat L&=&\sqrt{10}(d^\dag\times\tilde{d})^{(1)},\\
\label{Q}
\hat Q&=& s^{\dagger}\tilde d
+d^\dagger\tilde s+\chi(d^\dagger\times\tilde d)^{(2)}.
\end{eqnarray} 
The symbol $\cdot$ represents the scalar product. Here the
scalar product of two operators with angular momentum $L$ is defined as 
$\hat T_L\cdot \hat T_L=\sum_M (-1)^M \hat T_{LM}\hat T_{L-M}$ where 
$\hat T_{LM}$ corresponds to the $M$ component of the operator 
$\hat T_{L}$. The operator 
$\tilde\gamma_{\ell m}=(-1)^{m}\gamma_{\ell -m}$ (where $\gamma$
refers to $s$ and $d$ bosons) is introduced to ensure the
tensorial character under spatial rotations. Note that in realistic
calculations $\epsilon_d>0$ and  $\kappa>0$ \cite{Cast88}. 
It is common in this
approach to use for the $E2$ transition operator the form 
\begin{equation}
\label{te2}
\hat T(E2)=q_{\mbox{eff}} \hat Q  ,
\end{equation} 
in which $q_{\mbox{eff}}$ denotes the effective charge and 
$\hat Q$ has the same structure as in the Hamiltonian (\ref{Q}). 
This approximation is the basis of the so-called consistent-$Q$
formalism (CQF) \cite{Warn82}.

The Hamiltonian (\ref{ham1}) generates the spectrum of the
nucleus and in the following will be called ``local
Hamiltonian''. An extra part can be added to this Hamiltonian
(\ref{ham1}) without affecting the spectrum, that will be called
``global Hamiltonian''.

In the description of $BE$ using the IBM one has 
to distinguish between two
contributions: a global (rather big) part that corresponds to the 
bulk energy of the atomic nucleus, should 
change slowly ($BE^{gl}$) and comes from the ``global Hamiltonian'', 
and a local 
(rather small) part coming from the specific
structure of the given nuclei ($BE^{lo}$), {\it i.e.}~coming from the
``local Hamiltonian''. 
However, the global part has to be added {\it ad hoc}
using a given prescription that will be presented in this section. 
The simplest interpretation of the IBM global part comes from the LDM,
and somehow, both contributions must be related. For the 
different series of isotopes they result into a quadratic 
behavior in $BE$ and a linear one in $S_{2n}$. This IBM
description resembles the Strutinsky 
method \cite{Stru67,Stru68} in the sense that a part
(global part) takes care of the main part of the $BE$, while a
second part (local part) modifies this bulk $BE$ and generates 
the spectrum of the nucleus.

\subsection{The global part of the {\em BE} ($S_{2n}$) in the IBM}
\label{sec-ibm-be-global}
The global part of the $BE$ in the IBM ($BE^{gl}$)
comes from that part of the Hamiltonian that does not affect 
the internal excitation energies. 
Those
terms are related with the Casimir operators of $U(6)$, 
{\it i.e.}~$\hat C_1[U(6)]$ and $\hat C_2[U(6)]$ 
and can be written in terms of the total number of bosons, 
$\hat N$,
\begin{equation}
\label{ham-gl}
\hat H^{gl}=-E_0-{\cal A}-{{\cal B}\over 2} \hat{N}(\hat{N}-1). 
\end{equation} 
Its contribution to the binding energy reads as,
\begin{equation}          
\label{be-gl-ibm}
BE^{gl}(N)=E_0+{\cal A} N + {{\cal B}\over 2} N (N-1).
\end{equation} 
The corresponding contribution to $S_{2n}$ is linear in the
number of bosons:
\begin{equation}
\label{s2n-gl-ibm} 
S_{2n}^{gl}(N)=({\cal A}-{\cal B}/2)+{\cal B} N.
\end{equation}
In order to avoid ambiguities it is assumed in these expressions 
that $N$ always corresponds to the number of nucleons pairs, 
considered as particles and is never considered as holes. 
We come back to this delicate aspect in section
\ref{sec-mid}. 

In the latter expressions, it is implicit that the coefficients 
${\cal A}$, ${\cal B}$ and $E_0$ are constant for chains of isotopes 
(fixed $Z$) when 
the value of $N$ changes, except when crossing the mid-shell or
passing between major shells, {\it i.e.}~it provides a linear
contribution. However, this assumption is
not clear {\it a priori}. To find a mathematical proof of the 
constancy of ${\cal A}$ and ${\cal B}$  is a difficult 
task. However, one can find a number of arguments 
based on LDM, shell-model, and IBM itself, that imply such a
constancy.  
\begin{itemize}
\item A LDM argument: In section \ref{sec-ldm-be} we noticed
that the LDM gives a satisfactory global description of the $BE$
throughout the whole mass table. 
This description cannot reproduce fine details, but
it is able to explain the observed linear behavior of
$S_{2n}$ for series of isotopes. This behavior is the same 
as the one obtained
from (\ref{s2n-gl-ibm}), using ${\cal A}$ and ${\cal B}$ 
constants, and therefore supports our hypothesis. 
\item A shell-model argument: 
Another justification  is based on the
shell-model, in particular in the use of the modified 
surface-delta-interaction (MSDI). It is well known 
that the surface delta
interaction (SDI) gives a  good description of energy spectra 
although it also results in a number of systematic 
discrepancies with respect to the reproduction of the experimental levels.
This discrepancy is especially notable for nuclear binding energies.
It was shown that this description could be largely improved when
changing the position of the energy centroids for the multiplets with
different isospin. The modification of the
interaction gave rise to the MSDI \cite{Brus77}. 
The most important point for our present 
discussion is that this new element in the two-body 
interaction, if one keeps
the parameters constant, gives rise to a quadratic
dependence in the nuclear binding energy, 
equivalent to the one we obtain in eq.~(\ref{be-gl-ibm}). 
\item An IBM argument: A third justification comes from an IBM
analysis. It will be shown in sections \ref{sec-calc-ab} and
\ref{sec-real} that our {\it ansatz} provides
an extremely good description of $S_{2n}$ for chains of isotopes 
in the region from $Z=50$ to $Z=82.$    
\end{itemize}

\subsection{The local part of the $BE$ ($S_{2n}$) in the IBM: the
symmetry limits}
\label{sec-ibm-be-lim}

The local contribution to the $BE$ ($BE^{lo}$)
comes from the IBM Hamiltonian that
gives rise to the nuclear spectrum. This local contribution should be
added to the fully linear part presented in previous subsection.
A first approximation to this
Hamiltonian comes from studying  the symmetry limits of the model. 
Such limiting Hamiltonians do not correspond to realistic 
Hamiltonians but can be used as a  
good starting point. In the present discussion, the parameters of the 
different local Hamiltonians are kept constant, 
which is not a realistic
hypothesis for long chains of isotopes. Note that the global
Hamiltonian remains unchanged for the whole chain. Therefore, 
the following results will be applicable if we cut the chain of isotopes 
into smaller intervals and if we change the value of the parameters of
the local Hamiltonian only between intervals. 

The symmetry limits, called dynamical symmetries 
of the IBM, correspond to particular choices of the
Hamiltonian that give rise to analytic expressions for the energy
spectra (which is the reason of its usefulness). 
At the same time the eigenstates exhibit certain symmetries that
allow to classify them in a simple way. The symmetry limits appear
when the Hamiltonian is written in terms of a particular combination
of Casimir operators. Next we succinctly review the three cases 
that were discussed before \cite{Iach87}.   

\begin{itemize}
\item \underline{$U(5)$ limit.}

The local Hamiltonian that gives rise to the $U(5)$ symmetry limit
can be written as,
\begin{equation}
\label{h-u5}
\hat{H}_{U(5)} = \varepsilon\, \hat C_1[U(5)] + \alpha\, 
\hat C_2 [U(5)] + 
\beta\, \hat C_2 [O(5)] + \gamma\, \hat C_2 [O(3)], 
\end{equation} 
where $\hat C_n[G]$ stands for the Casimir operator of order $n$ of
the group $G$. The ground state of this Hamiltonian can be written as 
\begin{equation} 
\label{sta-u5}
| 0^+_{gs} \rangle = |[N], n_d=0, v=0, L=0 \rangle,
\end{equation} 
where $[N]$, $n_d$, $v$, and $L$ are the appropriate labels that
completely specify an eigenstate of the Hamiltonian (\ref{h-u5}) (see
{\it e.g.}~\cite{Iach87,Fran94}). 
The eigenvalue of (\ref{h-u5}) for a general state is obtained as,
\begin{equation} 
\label{ener-u5}
E_{U(5)}=\varepsilon\, n_d + \alpha\, n_d (n_d+4) + \beta\, v (v+3)
+\gamma\, L (L+1).
\end{equation} 
As a consequence $BE_{U(5)}$=0 and $S_{2n}^{U(5)}$=0. It is clear 
that there is no local contribution to the $BE$ in the case of the 
$U(5)$ limit.

\item \underline{$SU(3)$ limit.}

In the case of the $SU(3)$ dynamical symmetry, the local 
Hamiltonian reads as, 
\begin{equation} 
\label{h-su3}
\hat{H}_{SU(3)} = \delta\, \hat C_2[SU(3)] + \gamma\, \hat C_2[O(3)].
\end{equation} 
The ground state for this Hamiltonian corresponds to,
\begin{equation} 
\label{sta-su3}
|0^+_{gs} \rangle =|[N], (\lambda=2N,\mu=0), \kappa=0, L=0 \rangle, 
\end{equation} 
where $[N]$, $(\lambda,\mu)$, $\kappa$, and $L$ are the appropriate 
labels for completely specifying an eigenstate of the Hamiltonian 
(\ref{h-su3}) (see {\it e.g.}~\cite{Iach87,Fran94}). 
The eigenvalues corresponding to the Hamiltonian (\ref{h-su3}), 
for a general state, can be written as,
\begin{equation} 
\label{ener-su3}
E_{SU(3)}=\delta\, (\lambda^2 +\mu^2 +\lambda\mu+3\lambda+3\mu)+
\gamma \,L(L+1).
\end{equation} 
In this case the binding energy results into the expression,
\begin{equation}
\label{be-su3}
BE_{SU(3)}=-\delta (4N^2+6N).
\end{equation} 
The value of $S_{2n}$ for particles becomes,  
\begin{equation} 
\label{s2n-su3-p}
S_{2n}^{SU(3)}=-\delta (8N+2),
\end{equation} 
while for holes this becomes 
\begin{equation} 
\label{s2n-su3-h}
S_{2n}^{SU(3)}=\delta (8N+10),
\end{equation} 
where $\delta<0$ in realistic calculations. It should be noted 
that the local contribution to $S_{2n}$ is also linear in the number
of bosons. This contribution should be added to the global part of the
Hamiltonian.

\item \underline{$O(6)$ limit.}

The $O(6)$ symmetry limit corresponds to the following Hamiltonian,
\begin{equation} 
\label{h-o6}
\hat{H}_{O(6)} = \zeta\,\hat C_2[O(6)]+ \beta\,\hat C_2[O(5)]+ 
\gamma\,\hat C_2[O(3)].
\end{equation} 
The ground state for this Hamiltonian reads as,
\begin{equation} 
\label{sta-o6}
| 0^+_{gs} \rangle = |[N],\sigma=N, \tau=0, L=0 \rangle,
\end{equation} 
where $[N]$, $\sigma$, $\tau$, and $L$ completely characterise an
eigenstate of (\ref{h-o6}) (see {\it e.g.}~\cite{Iach87,Fran94}). 
The energy eigenvalues of the Hamiltonian
(\ref{h-o6}) can be written as,
\begin{equation} 
\label{ener-o6}
E_{O(6)}=\zeta\,\sigma(\sigma+4)+ \beta\,\tau(\tau+3) + 
\gamma\,L(L+1).
\end{equation} 
In this case the binding energy results into the expression,
\begin{equation}
\label{be-o6}
BE_{O(6)}=-\zeta (N^2+4N).
\end{equation} 
The value of $S_{2n}$ for particles becomes,
\begin{equation} 
\label{s2n-o6-p}
S_{2n}^{O(6)}=\zeta (2N+3),
\end{equation} 
while for holes it reads,
\begin{equation} 
\label{s2n-o6-h}
S_{2n}^{O(6)}=-\zeta (2N+5).
\end{equation} 
In the more realistic calculations $\zeta<0$. It should be noted that 
the local contribution to $S_{2n}$ is again linear in the number of 
bosons. This contribution should be added to the global part of the
Hamiltonian.
\end{itemize}

At this point, it should become clear that the local
IBM Hamiltonian, corresponding to the three dynamical symmetries,
does not change the linear behavior of $S_{2n}$ (coming from the
global part). In the
case of the $U(5)$ limit there is no extra contribution to $S_{2n}$,
while for the $SU(3)$ and $O(6)$ limits only a change in the values of
the slope and the intercept of $S_{2n}$ is introduced. 
The analysis should only be valid within the smaller intervals and thus
non-linear behavior in $S_{2n}$ could appear if the character of the
nuclei, along a series of isotopes, is changing  from one symmetry
limit into another one. An extra source for deviations of a linear 
behavior arises when the 
parameters of the local Hamiltonian themselves do change from one 
nucleus to another 
nucleus, even preserving the dynamical symmetry. Note that the global
contribution remains linear. 

\subsection{The local part of the {\em BE} ($S_{2n}$) in the IBM: 
  near the symmetry limits}
\label{sec-ibm-be-near-lim}
In this subsection we complete the previous analysis, but now we  
study more complex situations albeit still in an analytical
approximation. This will form a good
starting point in order to carry out  a complete numerical 
analysis of $S_{2n}$ using the IBM. 

Here, we consider the Hamiltonian (\ref{ham1}), which will prove to be
extremely useful for our purpose. This 
Hamiltonian encompasses the three symmetry limits for particular 
choices of the parameters and the so called
transitional regions. The transitional regions are intermediate 
situations between the symmetry limits, where one observes rapid 
structural changes in the nuclei. One can identify three different
transitional regions: (a) structural changes between
spherical ($U(5)$) and well deformed nuclei ($SU(3)$); (b)
structural changes
from spherical ($U(5)$) to $\gamma$-unstable nuclei ($O(6)$) and (c)
structural changes from well-deformed ($SU(3)$) 
to $\gamma$-unstable nuclei ($O(6)$). 
One observes that the borders of the transitional regions 
correspond to the dynamical symmetries (indicated between
parenthesis).

The idea here is to consider the main part of the local Hamiltonian  
corresponding to a given symmetry limit plus a small correction
term that allows us to explore the transitional region and that 
can be treated using perturbation theory. Next, we discuss three 
different situations depending on the main part of the local
Hamiltonian. 
\begin{itemize}
\item \underline{Near the $U(5)$ limit.}

The vibrational limit appears when $\kappa=0$ in Hamiltonian
(\ref{ham1}). If $\kappa\neq 0$ the wave function (\ref{sta-u5}) 
is only an approximate solution to the problem if 
$|\epsilon_d|>>|\kappa|$. The result is in principle independent of  
$\chi$, but performing a simple analysis, one notices that the range of
applicability of the results depends on $\chi$. So, if one 
numerically diagonalizes the Hamiltonian (\ref{ham1}) for $N=8$ and
for different values of $\kappa/\epsilon_d$, one obtains that, 
even for a ratio equal to $0.03$ and with $\chi=0$, the overlap
$\langle gs |[N], n_d=0, v=0, L=0 \rangle$ is equal to $0.913$.
For $\chi=-1$ this overlap equals $0.876$, and for 
$\chi=-\sqrt{7}/2$ it equals $0.833$. So it becomes clear that 
the range of applicability is quite
narrow and even diminishes when $|\chi|$ increases.  

In the following discussion we assume $\kappa'=0$ because its 
contribution to the $BE$ always vanishes. If one calculates the mean 
value of (\ref{ham1}), using the eigenstates (\ref{sta-u5}), the  
result becomes,
\begin{eqnarray} 
\label{near-u5-be1}
BE&=&-\langle 0^+_{gs-U(5)}|  
\epsilon_d \hat{n}_d - \kappa \hat{Q}\cdot\hat{Q}
| 0^+_{gs-U(5)} \rangle\nonumber\\
&=&\kappa\langle 0^+_{gs-U(5)}|\hat{Q}\cdot\hat{Q}
|0^+_{gs-U(5)}\rangle.
\end{eqnarray} 
The first term vanishes because $n_d=0$ in the $U(5)$ ground state
(see eqs.~(\ref{sta-u5}) and (\ref{ener-u5})). 
In order to calculate the remaining part, we consider
the expression of the quadrupole operator (\ref{Q}) and we take
into account that every $\tilde d$ operator acting directly 
on the {\it ket} state, or every $d^\dag$ operator acting directly 
on the {\it bra} state, gives a vanishing contribution. 
The $BE$ result then becomes,
\begin{equation}   
\label{near-u5-be2}
BE=5\kappa N.
\end{equation} 
The two-neutron separation energy for particles can be written as,
\begin{equation} 
\label{near-u5-s2n-p}
S_{2n}= 5\kappa,
\end{equation} 
while for holes it becomes,
\begin{equation} 
\label{near-u5-s2n-h}
S_{2n}=-5\kappa.
\end{equation} 
As a consequence, near the vibrational limit, the local Hamiltonian
only gives a constant contribution to $S_{2n}$.

\item \underline{Near the $SU(3)$ limit.}

A particular case of a rotational nucleus corresponds to the $SU(3)$
limit. In this case $\epsilon_d=0$ and $\chi=-\sqrt{7}/2$ are the
parameters that show up in the
Hamiltonian (\ref{ham1}). If we include $\epsilon_d\neq 0$ such that
$|\kappa|>>|\epsilon_d|$, the wave function (\ref{sta-su3}) will still
be a good approximation to the exact 
solution. In order to explore up to which values of 
$\epsilon_d$ one can use perturbation theory, we calculate 
the overlap between the state
(\ref{sta-su3}) and the exact solution for a system with $N=8$
bosons. The result we obtain is that, even for a ratio
$|\epsilon_d|/|\kappa|=10$ ,the overlap is larger than $0.9$ (in
particular equal to $0.942$). So we can use the present 
approximation in regions quite far from the $SU(3)$ symmetry limit.

By calculating the expectation value of
(\ref{ham1}), using the state (\ref{sta-su3}), the binding energy
becomes, 
\begin{eqnarray} 
\label{near-su3-be1}
BE&=&-\langle 0^+_{gs-SU(3)}|  
\epsilon_d \hat{n}_d - 
\kappa \hat{Q}^{\chi=-\sqrt{7}/2}\cdot\hat{Q}^{\chi=-\sqrt{7}/2}
| 0^+_{gs-SU(3)} \rangle\nonumber\\
&=&-\epsilon_d\langle 0^+_{gs-SU(3)}|\hat{n}_d| 0^+_{gs-SU(3)} \rangle 
\nonumber\\
&~&+\kappa\langle 0^+_{gs-SU(3)}|
\hat{Q}^{\chi=-\sqrt{7}/2}\cdot\hat{Q}^{\chi=-\sqrt{7}/2}
|0^+_{gs-SU(3)}\rangle.
\end{eqnarray}
The expectation value of $\hat n_d$ in the $SU(3)$ limit is known 
\cite{Iach87}, with as a result,
\begin{equation} 
\label{nd-su3}
\langle 0^+_{gs-SU(3)}|\hat{n}_d| 0^+_{gs-SU(3)} \rangle=
{4N(N-1)\over 3(2N-1)}.
\end{equation} 
On the other hand 
$\hat{Q}^{\chi=-\sqrt{7}/2}\cdot\hat{Q}^{\chi=-\sqrt{7}/2}$ 
is directly related with
the $SU(3)$ Casimir operator appearing in eq.~(\ref{h-su3}),
\begin{equation}
\label{qq-c-su3}
\hat{Q}^{\chi=-\sqrt{7}/2}\cdot\hat{Q}^{\chi=-\sqrt{7}/2}
={1\over 2}\hat C_2[SU(3)]-
{3\over 8}\hat{L}\cdot\hat{L}.
\end{equation} 
Finally, one obtains the result,
\begin{equation}
\label{near-su3-be2}
BE=-\epsilon_d{4N(N-1)\over 3(2N-1)}+\kappa (2N^2+3N).
\end{equation}

The final expression for $S_{2n}$ in the case of particles becomes, 
\begin{equation} 
\label{near-su3-s2n-p}
S_{2n}=-\epsilon_d{8(N-1)^2\over 3(4N^2-8N+3)}+\kappa(4N+1),
\end{equation} 
while in the case of holes it reads,
\begin{equation} 
\label{near-su3-s2n-h}
S_{2n}=\epsilon_d{8N^2\over 3-12 N^2}-\kappa(4N+5).
\end{equation} 
The first term in both eqs.~(\ref{near-su3-s2n-p}) and
(\ref{near-su3-s2n-h}), formally introduces a quadratic $N$
dependence. However, studying the expression for $S_{2}$ in more
detail, one observes that the final result is almost $N$ independent. 
In the case of 
$N\rightarrow\infty$, the asymptotic value is $0.667$. Already for
$N=8$ one obtains the value $0.670$ in the case of particles and 
$0.669$ in the case of holes. 
As a conclusion, the situation close to the
$SU(3)$ limit also gives rise to a linear behavior in $S_{2n}$.

\item \underline{Near the $O(6)$ limit.}

The $\gamma$-unstable nuclei are well described using the $O(6)$
limit. In this case one should make the choice $\epsilon_d=0$ 
and $\chi=0$ in the
Hamiltonian (\ref{ham1}). If we include $\epsilon_d\neq 0$ such that
$|\kappa|>>|\epsilon_d|$, the wave function (\ref{sta-o6}) becomes a  
good approximation to the exact solution. In order to find out 
how far can one proceed in the choice of the value of $\epsilon_d$, 
we calculate the overlap between the state
(\ref{sta-o6}) and the exact solution for a system with $N=8$
bosons. The result is such that, even for a ratio
$|\epsilon_d|/|\kappa|=10$, the overlap is larger than $0.9$ (in
particular equal to $0.916$).

The calculation of the expectation value of
(\ref{ham1}), using the eigenstate (\ref{sta-o6}), results in the
expression,
\begin{eqnarray} 
\label{near-o6-be1}
BE&=&-\langle 0^+_{gs-O(6)}|  
\epsilon_d \hat{n}_d - 
\kappa \hat{Q}^{\chi=0}\cdot\hat{Q}^{\chi=0}
| 0^+_{gs-O(6)} \rangle\nonumber\\
&=&-\epsilon_d\langle 0^+_{gs-O(6)}|\hat{n}_d| 0^+_{gs-O(6)} \rangle 
\nonumber\\
&~&
+\kappa\langle 0^+_{gs-O(6)}|\hat{Q}^{\chi=0}\cdot\hat{Q}^{\chi=0} 
|0^+_{gs-O(6)}\rangle.
\end{eqnarray}
The expectation value of $\hat n_d$ in the $O(6)$ limit is known
\cite{Iach87}, with as a result,
\begin{equation} 
\label{nd-o6}
\langle 0^+_{gs-O(6)}|\hat{n}_d| 0^+_{gs-O(6)} \rangle=
{N(N-1)\over 2(N+1)}.
\end{equation} 
On the other hand $\hat{Q}^{\chi=0}\cdot\hat{Q}^{\chi=0}$ 
is directly related with
the $O(6)$ and $O(5)$ Casimir operators appearing in eq.~(\ref{h-o6}),
\begin{equation}
\label{qq-c-o6}
\hat{Q}^{\chi=0}\cdot\hat{Q}^{\chi=0}=\hat C_2[O(6)]-\hat C_2[O(5)].
\end{equation} 
Finally, the binding energy becomes,
\begin{equation}
\label{near-o6-be2}
BE=-\epsilon_d{N(N-1)\over 2(N+1)}+
\kappa (N^2+4N).
\end{equation} 
The value of $S_{2n}$ for particles results in the expression,
\begin{equation} 
\label{near-o6-s2n-p}
S_{2n}=-\epsilon_d{N^2+N-2\over 2(N^2+N)}+\kappa(2N+3),
\end{equation} 
and for holes it reads
\begin{equation} 
\label{near-o6-s2n-h}
S_{2n}=\epsilon_d{N(N+3)\over 2(N+1)(N+2)}-\kappa(2N+5).
\end{equation}
Again, the first term introduces a formal quadratic $N$
dependence. However, studying the expressions (\ref{near-o6-s2n-p})
and (\ref{near-o6-s2n-h}) in more detail, one observes that the $N$
dependence almost cancels.  In the case of 
$N\rightarrow\infty$ the asymptotic value is $0.5$. Already for
$N=8$ one reaches the value $0.486$ in the case of particles and 
$0.489$ in the case of holes. As a conclusion, the 
situations close to the $O(6))$ limit also give rise to a 
linear behavior in $S_{2n}$.
\end{itemize}
Note that the transitional region $SU(3)-O(6)$ cannot be treated using
the Hamiltonian (\ref{ham1}); a treatment based on perturbation
theory does not result into a closed expression for the binding energy. 

The results obtained in this subsection exhibit the same
characteristic as the ones obtained in section 
\ref{sec-ibm-be-lim}, in the sense that all
situations that have been analysed always give rise to a 
linear contribution 
in $S_{2n}$. The only way of obtaining deviations from a linear 
behavior is through the presence of systematic 
changes of the parameters in the local 
Hamiltonian. This approach has
been explored in detail in \cite{Garc01a} and gives rise, indeed, to
non-linearities in $S_{2n}$ for the transitional $U(5)-SU(3)$
and $U(5)-O(6)$ regions. 

Now that we have carried out the various schematic analyses of 
$S_{2n}$ in the IBM, we present a more 
realistic description of binding energies and $S_{2n}$ values in the
next subsection.

\subsection{Crossing the mid-shell}
\label{sec-mid}
In the previous subsection we have derived closed 
expressions of $S_{2n}$ for
the case of particles and for the case of holes, independently. 
However, the counting
of particles/holes produces some inconsistencies and problems: using the
expressions (\ref{be-gl-ibm}) and (\ref{s2n-gl-ibm}), the sign of
${\cal B}$
should be changed when crossing the mid-shell. On the other hand, 
when plotting the binding energies (or $S_{2n}$) in terms of $N$
(particles or holes) we obtain a ``function'' that is double 
valued and
can lead to some errors of interpretation. Moreover, 
we need two definitions of $S_{2n}$, one for particles and one for holes. 
A possible outcome of these inconsistencies is to introduce a new 
variable $\tilde N$ that represents the number of
valence particle pairs and that is related with the number of 
bosons $N$ (particles or holes) through the definition,
\begin{equation}
\label{n-tilde}
N=\left\{
\begin{array}{cc}
\tilde N& \mbox{for}~\tilde N\leq 
\frac{\displaystyle \Omega}{\displaystyle 2}\\
\Omega-\tilde N &\mbox{for}~\tilde N> 
\frac{\displaystyle \Omega}{\displaystyle 2}
\end{array}\right.,
\end{equation}  
where $\Omega=\sum (j +1/2)$ represents the size of the shell, 
that is the total number of bosons that can be put into that shell.
The value of the $BE$ then becomes,
\begin{equation}
\label{be-new}
BE(\tilde N)=E_0+{\cal A} \tilde N + 
{{\cal B}\over 2} \tilde N (\tilde N-1)+BE^{lo}_{IBM}(N(\tilde N)).
\end{equation} 
Using the variable $\tilde N$ we have a single definition 
of $S_{2n}$ for both, particles and holes, that reads,
\begin{equation} 
\label{s2n-new}
S_{2n}(\tilde N)=BE(\tilde N)-BE(\tilde N-1),
\end{equation}  
or, equivalently,
\begin{equation} 
\label{s2n-new2}
S_{2n}(\tilde N)=({\cal A}-{\cal B}/2) + {\cal B}\tilde N +
BE^{lo}_{IBM}(\tilde N)-BE^{lo}_{IBM}(\tilde N-1).
\end{equation}
Note that the expressions (\ref{s2n-su3-h}), (\ref{s2n-o6-h}), 
(\ref{near-su3-s2n-h}), and (\ref{near-o6-s2n-h}) can be used directly 
taking into account the relation (\ref{n-tilde}). 
The introduction of $\tilde N$ is just a formal trick, but it will 
simplify all further analysis. 

Sometimes it might be useful to represent $BE$ or 
$S_{2n}$ as a function of 
the atomic number $A$. In those cases it is trivial to rewrite the
equations (\ref{be-new}-\ref{s2n-new}) using $A$.

\begin{figure}
\begin{center}
\mbox{\epsfig{file=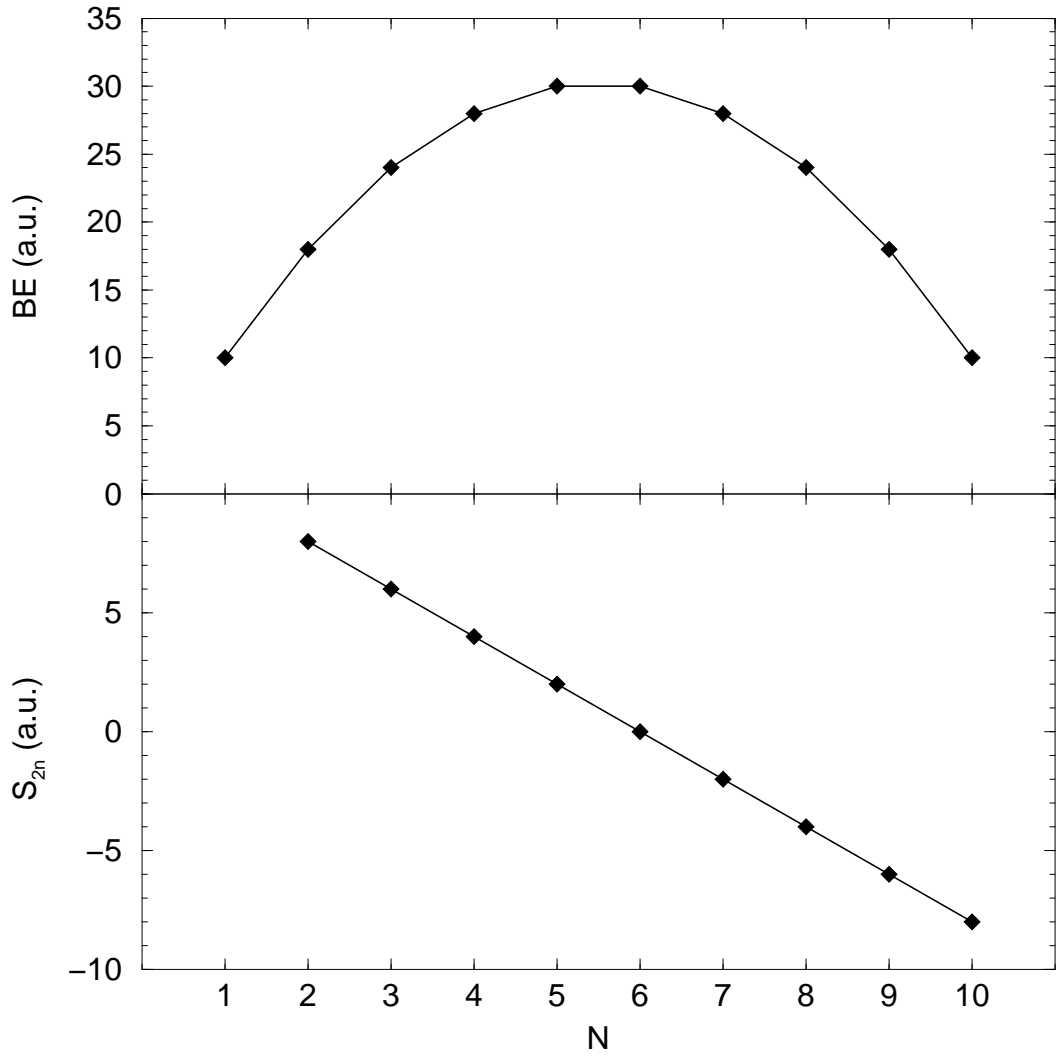,height=14.0cm,angle=-90}}
\end{center}
\caption{$BE$ (top) and $S_{2n}$ (bottom) for a pairing
  interaction in a single-$j$ shell with $\Omega=10$ and $G=1$ (in
  arbitrary units).}
\label{fig-pair-mid} 
\end{figure}

\begin{figure}
\begin{center}
\mbox{\epsfig{file=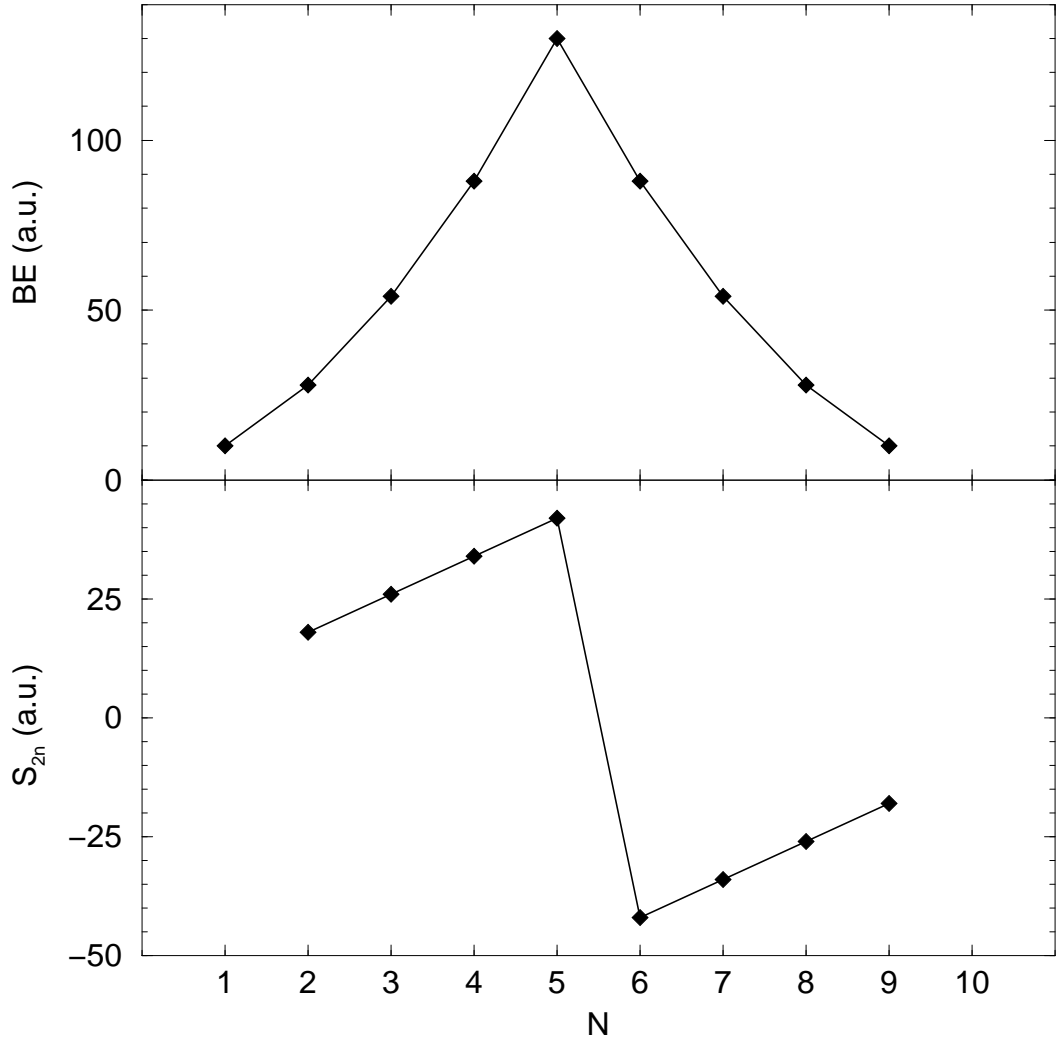,height=14.0cm,angle=-90}}
\end{center}
\caption{$BE$ (top) and $S_{2n}$ (bottom) for a $SU(3)$ IBM
  Hamiltonian, for $\Omega=10$ and $\delta=-1$ (in arbitrary units).}
\label{fig-su3-mid}
\end{figure}

Although with the introduction of $\tilde N$ we eliminate one
ambiguity of the IBM, there still appears a second 
problem that is intrinsic to the model. 
In order to illustrate it, we consider a
shell-model calculation. It is well known that, making the
appropriate changes in the shell-model Hamiltonian, 
{\it it does not matter} if one is using
particles or holes. Of course, changing from particles to holes when
crossing the mid-shell reduces considerably the size of
the model space. This freedom is intrinsic to the shell-model
because the Pauli principle avoids the over-counting of states within 
any shell. In the case of the IBM, the situation is completely different.
Working in a boson space, one can put an unlimited number of
bosons in a shell that has only room for $N_{max}=\Omega$
``bosons''.  This means that in the boson model space, the 
Pauli principle, or equivalently the size of
the ``boson shell'', is introduced by hand and, as a consequence, 
it is {\it obligatory} to change 
from particles to holes when crossing 
the mid-shell. The relevant point here is that this change induces a
discontinuity  in the value of $S_{2n}$ when crossing mid-shell. This
jump cannot represent a physical effect and  is not observed
experimentally either. In order to clarify
this point, we compare calculations using a pairing Hamiltonian 
in a single-$j$ shell in the fermion space, with the $SU(3)$ 
Hamiltonian (\ref{h-su3})\cite{Ring80,Heyd94}. 
In the case of pairing, the binding energy and $S_{2n}$ result as 
\begin{equation} 
\label{be-pairing}
BE=G\tilde N(\Omega-\tilde N+1),
\end{equation} 
and 
\begin{equation} 
\label{s2n-pairing}
S_{2n}=G(\Omega-2\tilde N+2),
\end{equation} 
respectively, where $G$ denotes the interaction strength  and 
$\Omega$ describes the size of the shell. 
In Fig.~\ref{fig-pair-mid}, we plot the expressions
(\ref{be-pairing}) and (\ref{s2n-pairing}) for
$G=1$ (in arbitrary units) and $\Omega=10$. 
One notices a smooth behavior
even when crossing the mid-shell. In the case of the $SU(3)$
Hamiltonian one has to use the equations (\ref{be-su3}),
(\ref{s2n-su3-p}), and (\ref{s2n-su3-h}). 
In Fig.~\ref{fig-su3-mid} we
plot these expressions for $\delta=-1$ (in arbitrary units) 
and $\Omega=10$. When comparing Fig.~\ref{fig-su3-mid} 
with Fig.~\ref{fig-pair-mid}, one observes 
clear differences with the pairing Hamiltonian, in
particular in the case of $S_{2n}$ (It should be noted that both cases
correspond to quite different physical situations. They are only used 
here in order to see the different behavior when crossing the mid-shell). 
There is a clear-cut unphysical
behavior when crossing the mid-shell. A solution to cope with this
inconsistency is obtained when we add the global part of $S_{2n}$ to the
local $S_{2n}$ term.
Thus, we keep an almost continuous variation in $S_{2n}$ by 
changing the value of 
${\cal A}$ and ${\cal B}$ when crossing the mid-shell. 
In the particular case of the $SU(3)$ limit we have just to change 
the value of ${\cal A}$ when crossing the
mid-shell in $8\Omega+12$, for $\delta=-1$ (in arbitrary units). 
In section \ref{sec-real} we shall
see that, even in realistic calculations, the solution to eliminate  
the spurious discontinuity in $S_{2n}$ is to change the 
parameters of the global part ($BE^{gl}$). It should be noted that 
in the case of odd $\Omega$, the equations (\ref{be-su3}) 
and (\ref{be-o6}) become invalid for
$N=\Omega/2+1/2$. For this value of $N$, the correct value 
is zero, while the expressions (\ref{be-su3}) and (\ref{be-o6}) 
give a value different from zero.  

\subsection{Deriving the global part: calculation of ${\cal A}$ 
and ${\cal B}$}
\label{sec-calc-ab}
Up to now we have carried out a schematic analysis of
the $S_{2n}$ observable, coming from the ``local Hamiltonian'' 
in the framework of the IBM.
We were able though to derive important conclusions. 
In this subsection we present a new approach for studying $S_{2n}$
values, spectra and transitions rates in a consistent way. 
In the next section,
we apply the method to nuclei belonging to the shell $Z=50-82$.
The method that will be used here was first discussed 
in \cite{Garc01a}. 

The key point of the method is the assumption that a linear global 
part ($S_{2n}^{gl}$), already presented in section
\ref{sec-ibm-be-global} 
needs to be added to the local contribution. 
Of course, the coefficients in this global contribution, 
${\cal A}$ and  ${\cal B}$, are taken as constant along the chain
of isotopes under study, except if crossing mid-shell or moving
between major shells (see previous subsection). 

In the previous subsections, the local Hamiltonian
was taken to correspond to a symmetry limit or to a situation close
to this. In the following discussion we consider more realistic
Hamiltonians. In principle, one has many possibilities 
for choosing a realistic Hamiltonian. 
The parameters of such a Hamiltonian should give rise to
a reasonable description of low-lying states of even-even
nuclei (energies and transition probabilities). 
However, in Ref.~\cite{Garc01a} it was shown that a 
correct description of spectroscopic properties does not 
always lead to a corresponding correct
description of the nuclear ground state properties, such as $S_{2n}$. 
A particular class of Hamiltonian that seems to provide good results 
for the excited states as well as for the ground state, is described
in Ref.~\cite{Chou97}. The Hamiltonian used corresponds to (\ref{ham1})
with $\kappa'=0$. The main characteristic is that the value of $\kappa$
is fixed for all even-even medium-mass and heavy nuclei to
$\kappa=0.030$~MeV. The values of $\epsilon_d$ and $\chi$ 
are adjusted to obtain the best possible description of energy spectra
and electromagnetic transition rates (the values of the parameters 
are given in Ref.~\cite{Chou97}). In this
framework the main observables that we intend to reproduce are
$E(2_1^+)$, $E(4_1^+)/E(2_1^+)$, $E(2_\gamma^+)$, 
$E(0_2^+)/(E(2_\gamma^+)-E(2_1^+))$, 
$B(E2;2_\gamma^+\rightarrow 0_1^+)/B(E2;2_\gamma^+\rightarrow 2_1^+)$,
and
$B(E2;2_\gamma^+\rightarrow 0_1^+)/B(E2;2_1^+\rightarrow 0_1^+)$
(where the label $\gamma$ refers to the $\gamma$ band, quasi-$\gamma$
band or even two-phonon-like band). In
the present paper we take as a guide the values of $\chi$ and
$\epsilon_d$ given in figures $10$ and $11$ of \cite{Chou97}, but 
the value of $\epsilon_d$ will be fine-tuned in order to obtain 
the best possible description for the energy spectra, but in all cases
the differences with respect to the values given in Ref.~\cite{Chou97}
are smaller than a $10\%$. It should be noted that the parametrisation
of the local Hamiltonian is fixed without considering the $S_{2n}$
values.

Once we have fixed the local IBM Hamiltonian (from
Ref.~\cite{Chou97}), it is a trivial task 
to deduce the linear part of $S_{2n}$ ($S_{2n}^{gl}$), {\it i.e.}
\begin{equation} 
\label{s2n-lin}
S_{2n}^{gl}\equiv {\cal A}+{\cal B} \tilde N=S_{2n}^{exp}-S_{2n}^{lo}.
\end{equation} 
(Note that for simplification we have made the substitution of 
${\cal A} -{\cal B}/2$ by ${\cal A}$). 
In practice the right hand side of eq.~(\ref{s2n-lin}) is not an exact 
relation but gives approximately a straight line. As a
consequence the linear part is derived from a best fit 
to the data points, 
obtained when plotting the right hand side in (\ref{s2n-lin}). 

It should be stressed that the values of ${\cal A}$ and ${\cal B}$  
thus obtained depend on the specific choice of the IBM local
Hamiltonian and 
as a consequence, for the best description of $S_{2n}$, one cannot mix 
local and global parts corresponding to different Hamiltonians. That
comes from the fact that different local Hamiltonians can give an
equally reasonable description of nuclear spectra, but their
contribution to the $BE$ value will be probably different, therefore,
in order to describe $S_{2n}$ correctly the global contribution 
(${\cal A}$ and ${\cal B}$) 
will be different. As a consequence the values
of ${\cal A}$ and ${\cal B}$ depend on the local Hamiltonian.

Although we already have a detailed prescription for extracting the
values of ${\cal A}$ and ${\cal B}$, we have to point out how to 
``operate'' when changing between major shells or when crossing the
mid-shell point. In principle the values of
${\cal A}$ and ${\cal B}$ will change 
when passing between different shells or 
crossing the mid-shell. That means that we have to 
consider different separate regions in our analysis.
\begin{itemize}
\item Moving between major shells:
In this case the values of ${\cal A}$ and ${\cal B}$ change, 
especially the value of ${\cal A}$ (the
intercept). In our calculations the nucleus corresponding to the
closed 
shell is excluded because the prescription that provides the
Hamiltonian (\ref{ham1}) is not applicable and intruder states 
become important in the description. Therefore, $S_{2n}$ values  
corresponding to a closed shell and to a closed shell plus two nucleons 
(of particles) will be excluded from the fit.  

\item Crossing the mid-shell:
As was explained previously, the IBM contains a clear deficiency when
crossing mid-shell because the Pauli principle is only included in
an approximate way. One of the main manifestation of
this fact appears when crossing mid-shell. We already pointed out
that a simple way to solve this artefact 
is to change the global (linear)
part of $S_{2n}$ for the second part of the shell. In all practical
cases we notice that no data points should be
excluded from the calculations as in the previous case. The mid-shell
point should be included in the calculation of $S_{2n}^{lo}$ 
before and after the mid-shell.
\end{itemize}

\subsection{Realistic calculations in the shell $50-82$}
\label{sec-real}
In the present section, we study the value of $S_{2n}$ for the
following chains of
isotopes: $^{114-144}_{\phantom{114-1}54}$Xe, 
$^{120-148}_{\phantom{114-1}56}$Ba, 
$^{124-152}_{\phantom{114-1}58}$Ce,
$^{128-154}_{\phantom{114-1}60}$Nd, 
$^{132-160}_{\phantom{114-1}62}$Sm,
$^{138-162}_{\phantom{114-1}64}$Gd, 
$^{148-166}_{\phantom{114-1}66}$Dy,
$^{150-168}_{\phantom{114-1}68}$Er,  
$^{152-178}_{\phantom{114-1}70}$Yb,
$^{158-184}_{\phantom{114-1}72}$Hf, 
$^{166-188}_{\phantom{114-1}74}$W,  
$^{170-196}_{\phantom{114-1}76}$Os, and
$^{176-200}_{\phantom{114-1}78}$Pt,  which are precisely
the isotopes analysed in Ref.~\cite{Chou97}. 
The superscripts refer to the
range of $A$ that we analyse in each series of isotopes.

The main aim of this section is not only to obtain a good
description of the experimental $S_{2n}$ 
\cite{Audi93,Audi95,Audi97}, 
but also to show that the
hypothesis of constancy for ${\cal A}$ and ${\cal B}$, {\it i.e.}~the
global part of the Hamiltonian gives a linear contribution, is 
fulfilled along a wide region of nuclei. Somehow both ends are
related.  

\begin{figure}
\begin{center}
\mbox{\epsfig{file=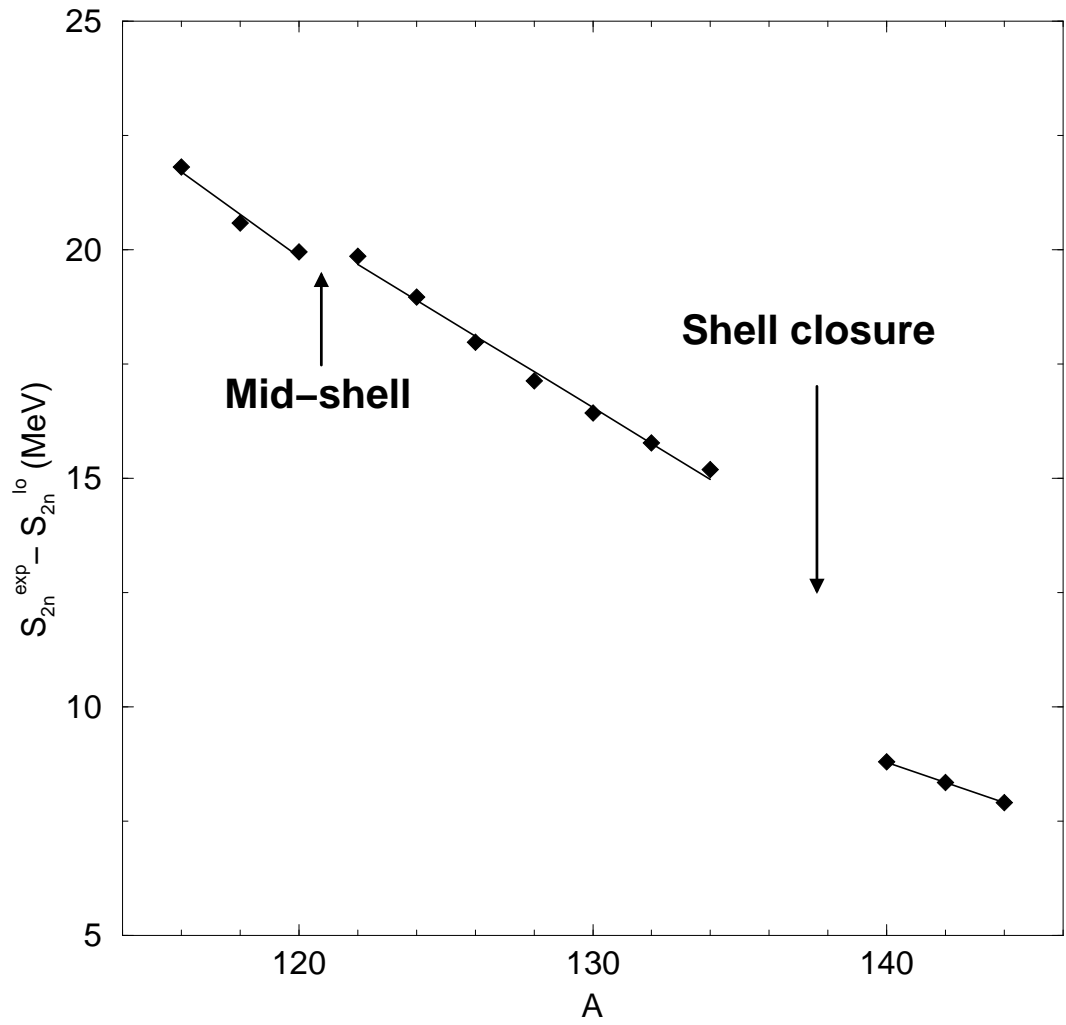,height=14.0cm,angle=-90}}
\end{center}
\caption{Differences $S_{2n}^{exp}-S_{2m}^{lo}$ (full diamonds)
  together with the regression line for Xe isotopes.}
\label{fig-xe-line}
\end{figure}

Following the prescription given in subsection \ref{sec-calc-ab}, we 
choose a realistic Hamiltonian for each series of isotopes and
we subtract the local contribution of $S_{2n}$,
$S_{2n}^{lo}$, from the experimental values, $S_{2n}^{exp}$. 
With these data we calculate the straight lines that give the best 
fits. We need to distinguish four regions: (a) 
$50<N\leq 66$, (b) $66<N<82$, (c) $84<N\leq 104$, and (d) $104<N<126$
(where $N$ represents the total number of neutrons). 
As can be observed, the points corresponding to the 
closed shell and to the closed shell plus two
neutrons are excluded, while the mid-shell point is taken into
account in the calculations. To illustrate the procedure in more
detail, we carefully explain the case of the Xe nuclei. 
In Fig.~\ref{fig-xe-line}  we show
the differences $S_{2n}^{exp}-S_{2n}^{lo}$ together with
the regression line. One observes 
three different regions: before mid-shell
in the shell $50-82$, after mid-shell for the same shell and before
mid-shell in the shell $82-126$. One notices the correctness of the
present description using a straight line for each region, separately.
The parameters of the Hamiltonian are given in table 
\ref{tab-xe-ham}; they correspond to the ones given in
\cite{Chou97}. The coefficients of the straight line in each region,
from the left to the right are, ${\cal A}=75.5\pm 9.9$ and 
${\cal B}=-0.464\pm 0.084$; ${\cal A}=67.4\pm 2.1$ and 
${\cal B}=-0.392\pm 0.016$; and ${\cal A}=39.90\pm 0.04$ 
and ${\cal B}=-0.2225\pm 0.0003$, respectively (all the
coefficients are given in MeV). Note that the
intercepts and slopes correspond to a representation
where we use the atomic number $A$ instead of the number of bosons 
$N$ or $\tilde N$. This criterion will be used along 
this whole subsection.
The error bar
derives from the standard deviation in obtaining the best fit 
and represents a measure of the goodness of the {\it ansatz}. 

\begin{figure}
\begin{center}
\mbox{\epsfig{file=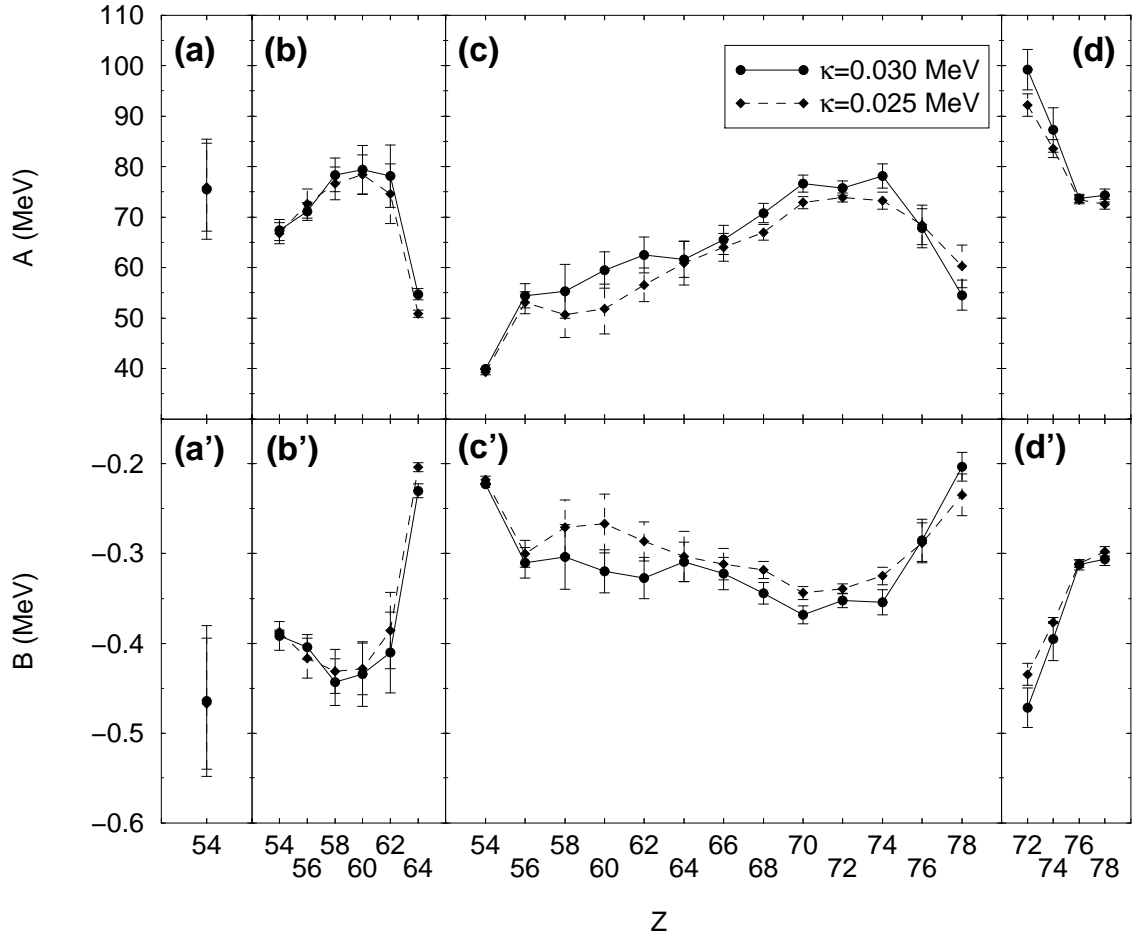,height=15.0cm,angle=-90}}
\end{center}
\caption{Values of ${\cal A}$ and ${\cal B}$ for different chains of
  isotopes (see text). Two alternatives calculation with different
  values of $\kappa$ are plotted.}
\label{fig-AB}
\end{figure}

We have carried out similar analyses for all the chains of isotopes
indicated in the beginning of this subsection, obtaining analogous
results. The parameters of the Hamiltonian that has been used are
given in reference \cite{Chou97}. In Fig.~\ref{fig-AB} we plot the
values of ${\cal A}$ and ${\cal B}$ for all the isotopes we studied. 
The panels (a)-(a'), (b)-(b'), (c)-(c'), and (d)-(d') correspond to 
the four regions defined previously. The error bars correspond to the
standard deviation in deriving ${\cal A}$ and ${\cal B}$. 
We also present
results from calculations using two different values of $\kappa$. 
This is done in order to show
the sensitivity of ${\cal A}$ and ${\cal B}$ with respect to 
small variations in the value of $\kappa$. In both sets of
calculations, the values of $\chi$ and $\epsilon_d$ are identical
(strictly speaking, due to the fine-tuning, $\epsilon_d$ is slightly 
different in both calculations). We can safely conclude that the values 
of ${\cal A}$ and ${\cal B}$ are not very sensitive to the 
choice of $\kappa,$ and the standard deviations are small enough 
to justify our {\it ansatz} that ${\cal A}$ and ${\cal B}$ are 
constant within the different mass regions. Note that the values of
${\cal A}$ and ${\cal B}$ change with $Z$ but remain constant for the
whole chain of isotopes (fixed $Z$), except when crossing the
mid-shell or changing of major shell. 
 
\begin{figure}
\begin{center}
\mbox{\epsfig{file=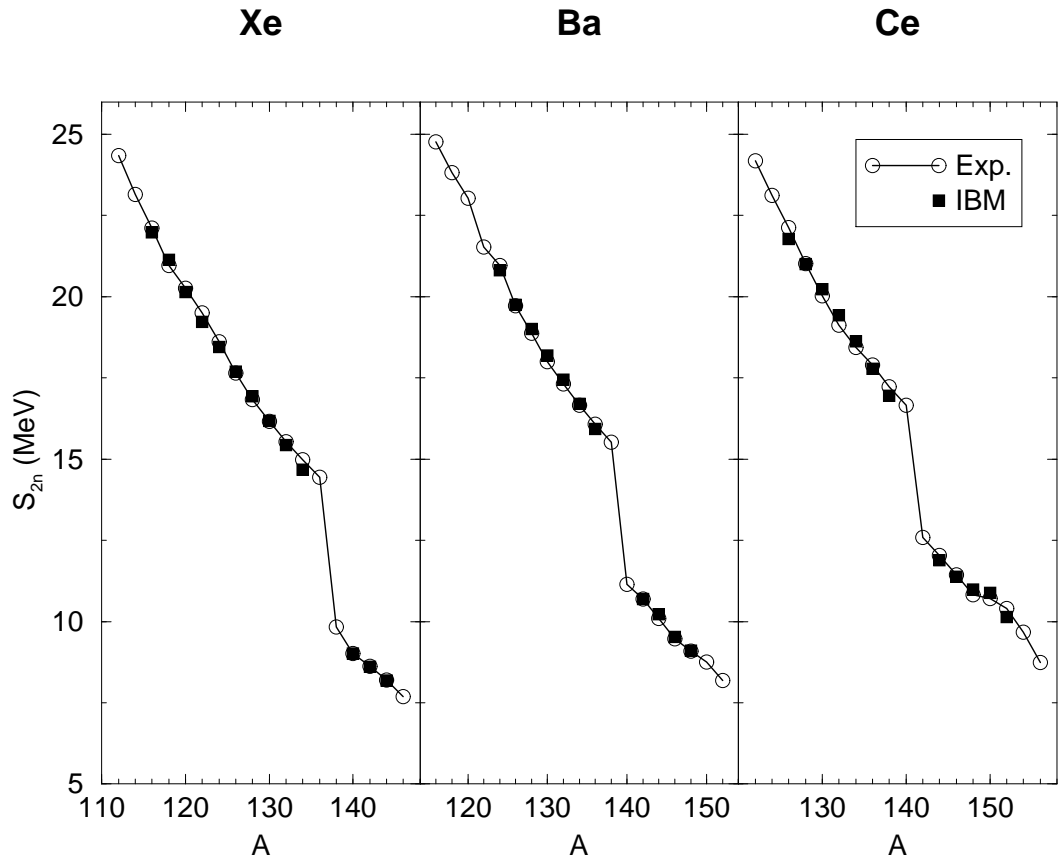,height=14.0cm,angle=-90}}
\end{center}
\caption{Comparison between the experimental $S_{2n}$ and the IBM 
  prediction for Xe, Ba, and Ce isotopes.}
\label{fig-s2n-xe-ba-ce}
\end{figure}

\begin{figure}
\begin{center}
\mbox{\epsfig{file=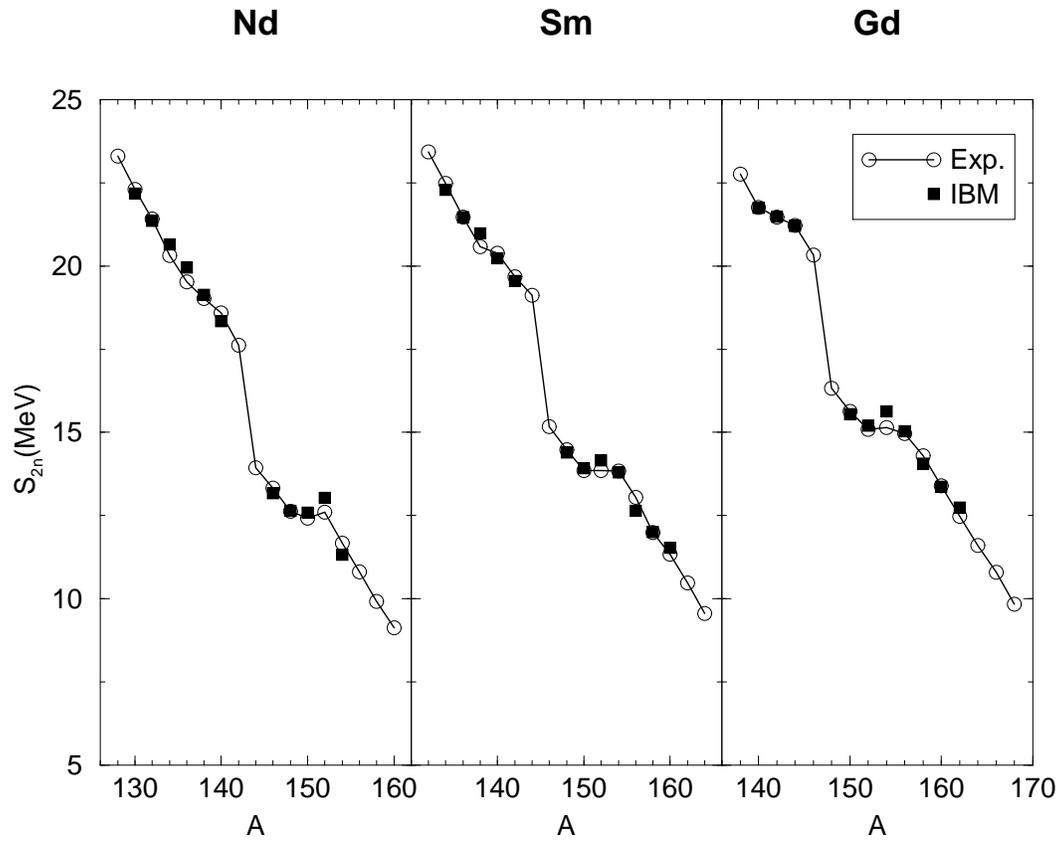,height=14.0cm,angle=-90}}
\end{center}
\caption{Comparison between the experimental $S_{2n}$ and the IBM 
  prediction for Nd, Sm, and Gd isotopes.}
\label{fig-s2n-nd-sm-gd}
\end{figure}

\begin{figure}
\begin{center}
\mbox{\epsfig{file=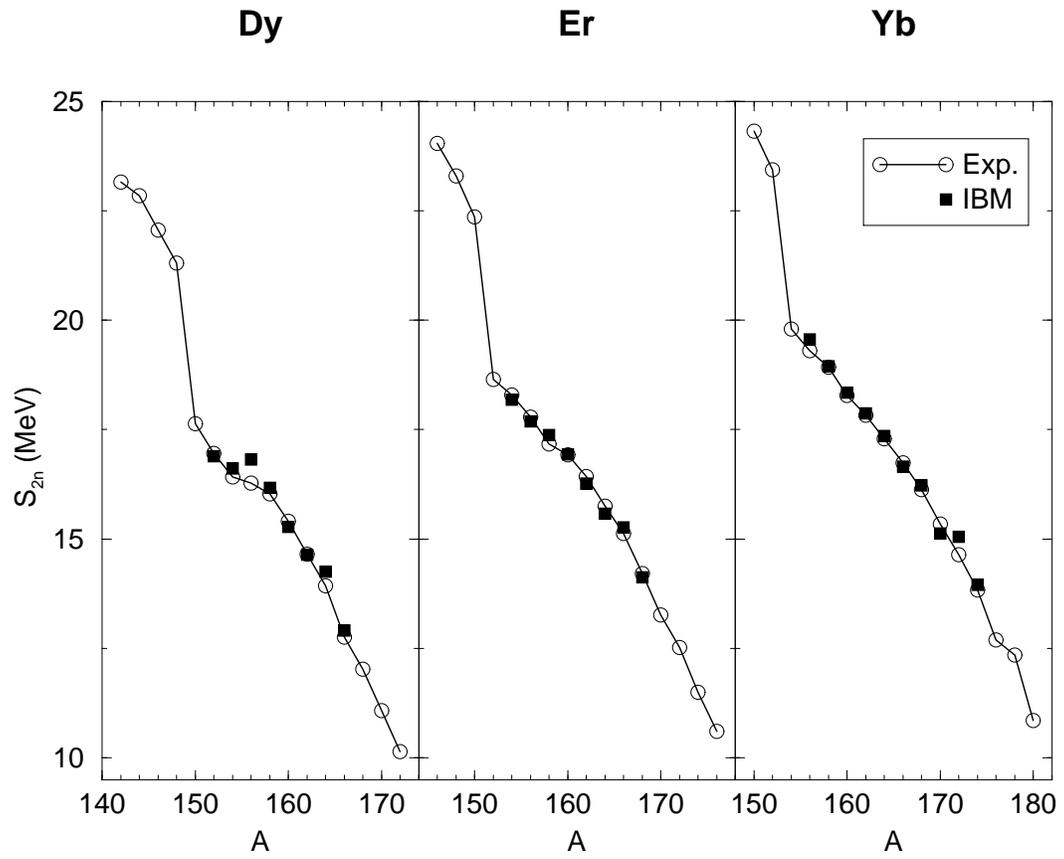,height=14.0cm,angle=-90}}
\end{center}
\caption{Comparison between the experimental $S_{2n}$ and the IBM 
  prediction for Dy, Er, and Yb isotopes.}
\label{fig-s2n-dy-er-yb}
\end{figure}

\begin{figure}
\begin{center}
\mbox{\epsfig{file=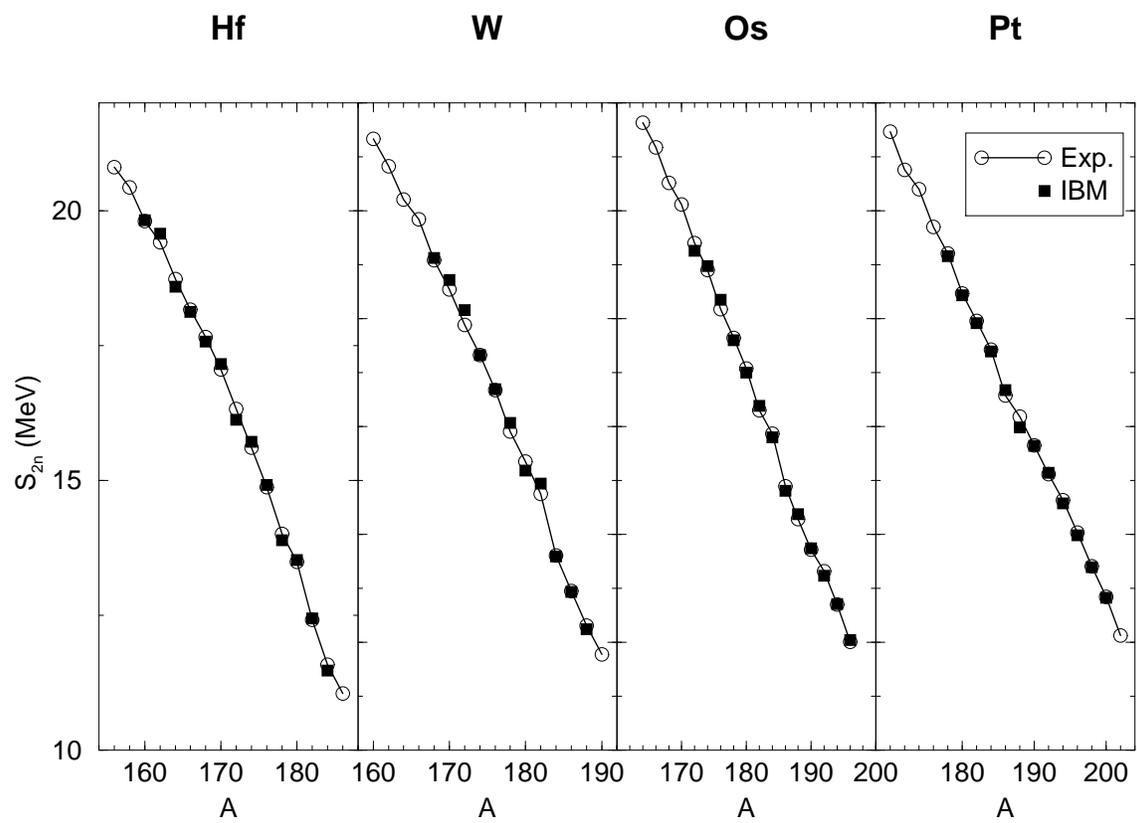,height=15.0cm,angle=-90}}
\end{center}
\caption{Comparison between the experimental $S_{2n}$ and the IBM 
  prediction for Hf, W, Os, and Pt isotopes.}
\label{fig-s2n-hf-w-os-pt}
\end{figure}

In figures \ref{fig-s2n-xe-ba-ce}, \ref{fig-s2n-nd-sm-gd},
\ref{fig-s2n-dy-er-yb}, and \ref{fig-s2n-hf-w-os-pt} we compared the 
experimental $S_{2n}$ values, $S_{2n}^{exp}$, with the
values predicted by the IBM, combining the linear and the local
part. In general, one obtains a rather good description, even in the
regions where the nuclear structure character is changing quite 
rapidly and important
deviations from the overall linear behavior appear. We stress that we
do reproduce $S_{2n}$ values and, at the same time, the
properties of the low-lying states in these nuclei. 
The IBM results correspond 
(figures \ref{fig-s2n-xe-ba-ce}-\ref{fig-s2n-hf-w-os-pt})  to
$\kappa=0.030$ MeV.

This analysis presented here, together with the results obtained 
in Ref.~\cite{Garc01a},
points towards an intimate relation between the correct reproduction 
of the nuclear excited states and
nuclear ground-state properties. A simultaneous description 
guarantees that the Hamiltonian that is used, is appropriate 
for the nuclei studied over a large mass region.

\section{The effect of intruder states: shape coexistence and shape
  mixing}
\label{sec-intru}
In the former sections, we have studied within various approaches
to nuclear structure: the liquid-drop approach (section
\ref{sec-ldm-be}), the nuclear shell-model (section \ref{sec-shell-be})
and the symmetry-truncated Interacting Boson Model (section
\ref{sec-ibm-be}). We have studied the behavior of an
important quantity, {\it i.e.}~$S_{2n}$, and we have tried to understand 
its variation over large regions of isotopes, in various mass
regions. A consistent set of conclusions follows from the above analyses.

\begin{figure}
\begin{center}
\mbox{\epsfig{file=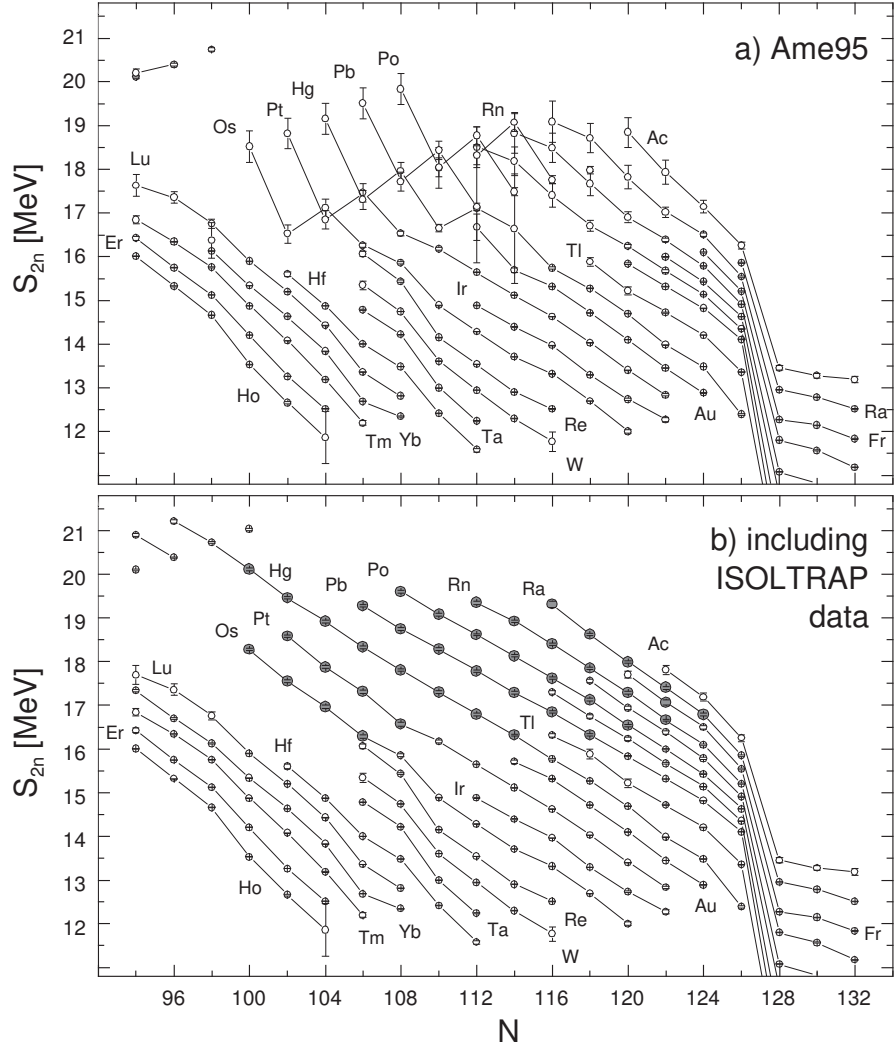,height=14.0cm,angle=0}}
\end{center}
\caption{Experimental two-neutron separation energies, $S_{2n}$, in
  the region of $Z=80$. a) Experimental values as determined from the
  analysis of Audi and Wapstra [43],
%  \cite{Audi93}, 
  including the update in 1995 [44]
%  \cite{Audi95}, 
  b) results obtained redoing the analyses of
  AME95, including the new ISOLTRAP data [20].
%  \cite{Schw01}.
  Full circles indicate $S_{2n}$ values that
  are either obtained for the first time or whose errors were
  decreased by at least a factor two.}  
\label{fig-s2n-exp-isol}
\end{figure}

It turns out, however, that if one starts looking to nuclear
masses with the highest possible precision
\cite{Boll96,Raim97,Schw98,Kohl99,Schw01,Beck00}
one becomes sensitive to 
localised correlations within the nuclear many-body system.
Such high-precision results in the lighter {\em sd}-shell 
region, at and very
near to neutron number $N=20$ for the Na, Mg nuclei, have brought evidence
for a new zone of deformation, albeit very localised in $(Z,N)$ 
values \cite{Warb90,Wood92}. Very recently, 
Bollen's group has succeeded in performing mass
measurements with the Penning trap mass spectrometer ISOLTRAP at 
ISOLDE/CERN in the neutron deficient region of the Hg, Pt, Pb, Po, Rn, 
and Ra nuclei \cite{Schw98,Kohl99,Schw01}. 
The results are discussed by Schwarz 
{\it et al.} \cite{Schw01} and are given in
Fig.~\ref{fig-s2n-exp-isol} and Figs.~\ref{fig-s2np-pb}, 
~\ref{fig-s2np-hg},
~\ref{fig-s2np-pt}, and ~\ref{fig-s2np-po}.  
Very particular effects in approaching the neutron
mid-shell region near $N=104$ show up.

A possible explanation might originate from the presence of shape
coexisting configurations in this particular mass region, which has been
discussed in much detail in \cite{Wood92,Heyd83}, and a
mixing between the intruding 
configuration and the ground-state, causing 
local deviations from a smooth linear trend. This is particularly 
striking in the Hg and Pt nuclei.

Because model spaces within the shell-model quickly become
prohibitively large if also particle-hole excitations across the
$Z=82$  closed proton shell are included, 
standard large-scale shell-model calculations cannot be carried 
out in a consistent way. Therefore, we 
discuss two approaches that might allow for such effects to be treated
in a consistent approximation: we start from the IBM 
but now taking into account $2p-2h$ excitations as the addition of 
two extra bosons (subsection \ref{sec-ibm-mixing}). We also study the 
modification of the nuclear ground-state binding energy as derived
from a macroscopic-microscopic study, 
in which potential energy surfaces (PES) 
are calculated taking into account competing shapes 
(spherical, prolate and oblate configurations) (subsection
\ref{sec-tpe}). 

\subsection{Local nuclear structure effects: intruder 
  excitations near closed shells within the Interacting Boson Model} 
\label{sec-ibm-mixing}
The effect of low-lying $0^{+}$ intruder excitations, which seems to
be related to $mp-nh$ excitations of nucleons across the adjacent 
closed shells, on energy spectra, electromagnetic properties, 
nuclear transfer data, etc., has been amply illustrated all through 
the nuclear mass table in the vicinity of closed shells
\cite{Wood92,Heyd83}. 
This holds in particular for heavy nuclei with the most
explicit examples in (and near to) the $Z=50$ (Sn) region near 
mid-shell at $N=66$ and in the $Z=82$ (Pb) region when approaching 
the mid-shell at $N=104$. A full study has been carried out by 
J.Wood {\it et al.} \cite{Wood92} which concentrates 
on the full mass table. 

\begin{figure}
\begin{center}
\mbox{\epsfig{file=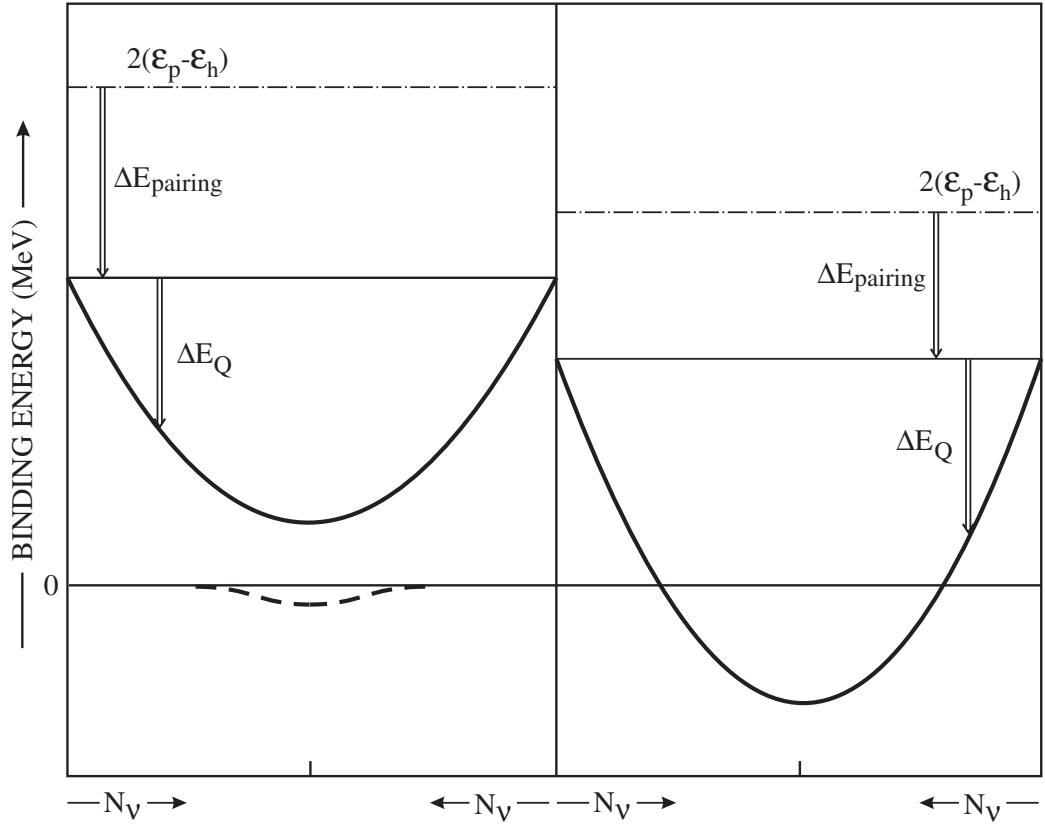,height=11.0cm,angle=0}}
\end{center}
\caption{Schematic representation of the effect of configuration
  mixing on the binding energy, plotting the different contributions 
  separately. On the left, it is assume that regular and intruder
  states seat far in energy. On the right, it is assume that the
  regular and intruder states cross.}
\label{fig-sch-mix}
\end{figure}

The inclusion of low-lying intruder states in even-even nuclei has
been modelled along the IBM by including an extra configuration with
two more bosons ($N+2$), that may interact with the regular 
configurations containing $N$ bosons 
\cite{Duva81,Duva82,Barf83,Hard97,Oros99,Cost99};
many calculations along these lines 
have been performed. Even though detailed 
calculations may well turn out to have a serious sensitivity to the 
choice of parameters describing the Hamiltonians
corresponding to the two subspaces \cite{Dele93a,Dele93b,Lehm97}, 
the general outcome remains very 
stable and gives the possibility to obtain 
(i) low-lying $0^{+}$ intruder states that
exhibit a very specific mass dependence, approximately 
described by the expression \cite{Heyd87}
\begin{equation} 
\label{ener-mix}
\Delta E_{Q} \simeq 2 \kappa \Delta N_{\pi}N_{\nu},
\end{equation}  
which expresses the extra binding energy that 
results from the interaction
of the extra proton pairs, $\Delta N_{\pi}$, 
with the valence neutron pairs,
$N_{\nu}$, using a quadrupole-quadrupole proton-neutron 
interaction with 
$\kappa$ as the force strength (see Fig.~\ref{fig-sch-mix} left), 
or, (ii) to come to a 
``crossing'' between the intruder configuration and the regular 
ground-state configuration (see Fig.~\ref{fig-sch-mix} right). The
latter effect causes a more deformed state to become the 
ground state and will subsequently show up in 
increased binding energy and, depending on the specific
nature of the intruder configuration, possibly gives rise to the 
appearance of a very
localised zone of deformation (island of inversion as called 
in the $N=20$ 
mass region \cite{Warb90,Wood92,Utsu99,Azai00,Caur98,Rodr00}). 
In both cases, local effects can 
cause the ground-state to exhibit very specific deviations 
from the otherwise mainly linear variation of $S_{2n}$. 
The former situation (i) will mainly appear when we are sitting in a
big 
shell like in the case in the Sn and Pb mass regions. The second
situation (ii), is more likely to show up near sub-shell closure 
($Z=40$, $N=58$, $Z=64$, $N=90$). This effect is depicted 
schematically in Fig.~\ref{fig-sch-mix} \cite{Heyd87}.

\begin{figure}
\begin{center}
\mbox{\epsfig{file=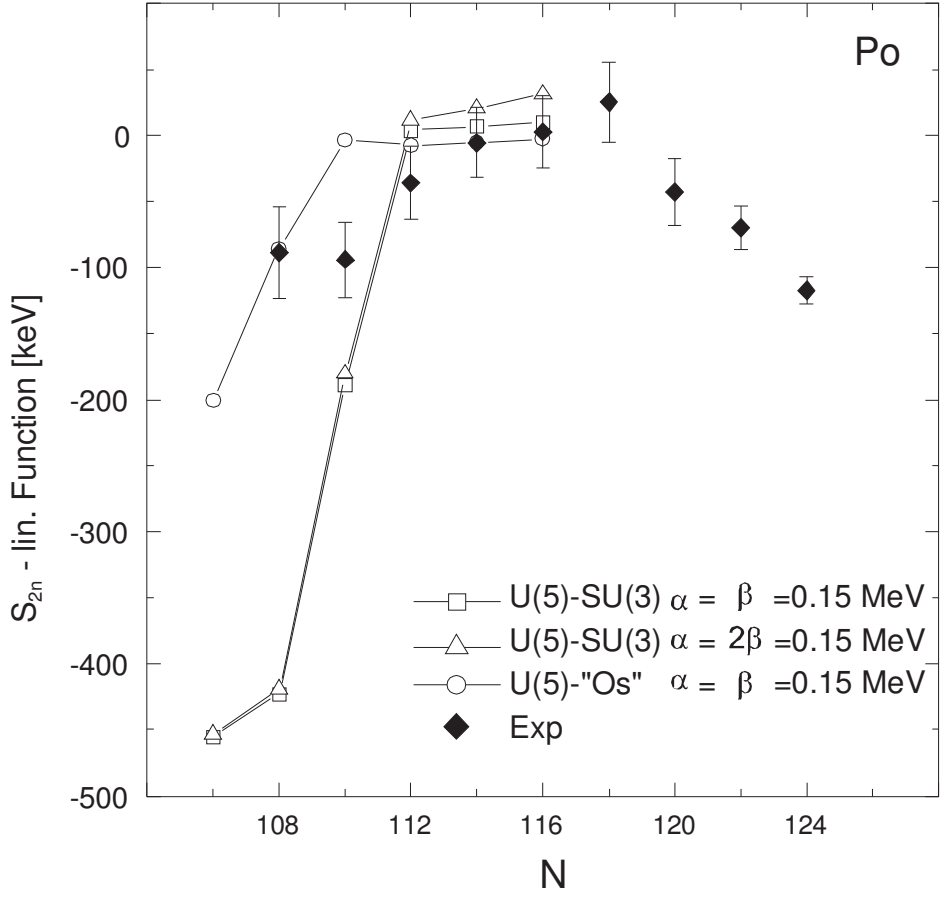,height=12.0cm,angle=0}}
\end{center}
\caption{Comparison of experimental $S_{2n}'$ values ($S_{2n}$ minus a 
  linear function) with the results of IBM configuration mixing 
  calculations for Po isotopes. Three different kinds of coupling 
  are considered: a 
  $U(5)-SU(3)$ dynamical symmetry coupling (open squares and
  triangles) and a more general IBM-1 coupling (including g bosons) 
  (open circles).}
\label{fig-mix-po}
\end{figure}

In the present discussion, we shall mainly concentrate on the Pb
region where an extensive data set has become 
available very recently (\cite{Wood92,Cost00} and references therein).  
We have carried out studies within the Interacting Boson Model approach
(IBM configuration mixing) in which low-lying intruder configurations 
are allowed to mix with the
regular ground-state configuration. Calculations have been carried out
for the Po isotopes with the aim of 
understanding the rapid lowering of
an excited $0^+$ state and the band on top of that
\cite{Oros99,Cost99}. Using a $U(5)-SU(3)$ dynamical symmetry 
coupling (ds) (using two different sets of coupling matrices) and also
a more general IBM-1 Hamiltonian for the intruder excitations (g)
\cite{Kibe94}, we
have studied the influence of mixing on the ground-state binding energy
and thus on the $S_{2n}$ values. One can see 
in Fig.~\ref{fig-mix-po} that the overall trend is
rather well reproduced and that in the lightest Po nucleus, where data
are obtained, albeit with a large error bar, a local drop of about $400$
keV to $150$ keV results, depending on how states are mixing 
(for more details see \cite{Cost99}).

In the Pb nuclei, no specific structure effects
outside of a linear variation in $S_{2n}$ show up except at the lowest
neutron number observed at present. This is consistent
with the excitation energy of the lowest $0^+$ intruder state not
dropping much below $0.8$ MeV \cite{Cost00}. 
The measured very slow $E0$ decay 
rates in the Pb nuclei \cite{Dupp87} are consistent with a
very weak mixing into the ground state and thus without local binding 
energy increase. 

Calculations in the Pt nuclei \cite{Hard97}, using similar methods, 
in the region where the two different families of states 
come close and interact with typical mixing matrix elements for 
the $0^+$ states of $100-200$ keV, result in 
a specific variation in the mass dependence of $S_{2n}$ values 
consistent with the observed data. 
Independent studies that have 
attempted to extract the mixing matrix element between the
ground-state and intruder-band
members, all come close to this value of $100-200$ keV as mixing matrix
element giving a consistent explanation \cite{Oros99,Cost99,Dupp90}. 

Even though it is not possible to derive every single
detail of the local $S_{2n}$ variations, all studies and the various
results on ground-to-intruder band state mixing  point towards the
interpretation that it is a localised interaction between the ground
state and the specific low-lying intruder $0^+$ states which is at the
origin of the observed effects. Moreover, there is a clear correlation
between the energy where the ground-state and intruder states have a
closest approach and the maximal deviation in $S_{2n}$ from a linear
variation.

\subsection{Macroscopic-microscopic calculations}
\label{sec-tpe}
Instead of describing binding energies from a shell-model approach
(standard large-scale shell-model calculations or the IBM approximation
in which the interactions amongst pairs form the central ingredients),
one can use a different method.
Here, we start from a model that combines the
macroscopic part of the total energy, with a microscopic part. This
latter part contains the  nuclear shell and pairing correlations near to the
Fermi level \cite{Ring80,Nils95,Stru67,Stru68,Brac72}. 
In shorthand notation, the total binding energy can be written as
\begin{equation}  
\label{be-strut}
BE = E_{LDM} + E_{shell} + E_{pair}.  
\end{equation}
The latter method is called the Strutinsky renormalization method and
has been applied in many mass regions 
(see {\it e.g.}~\cite{Moll81,Beng84,Moll81a,Moll88,Moll95,Moll97}).

The macroscopic part of the total energy was assumed to be given by a
Yukawa-plus-exponential mass formula of M\"oller and Nix
\cite{Moll81a,Moll88}. The shell correction was calculated using the
axially-deformed single-particle Woods-Saxon potential
\cite{Beng89,Cwio87}, with parameters as outlined if
Refs.~\cite{Cwio87,Dude82,Dude81}. This average potential has been
used previously to determine equilibrium shapes of coexisting
configurations \cite{Wood92} in the Pt-Ra region
\cite{Beng89,Beng87,Satu91,Naza93,May77}. The shell correction was
then calculated according to prescription as given by Brack {\it et
  al.}~\cite{Brac72},
\begin{equation} 
\label{stru-corr} 
E_{shell}  =  \sum_\nu e_{\nu}  -  
\int_{-\infty}^{\tilde{\lambda}} e \tilde{g} (e) \ de, 
\end{equation}
(with $e_{\nu}$ the single particle 
energies, $\tilde{g}(e)$ a smoothing
function and $\tilde{\lambda}$ the upper integration limit fixed by the
number of nucleons present).
For the residual particle-particle interaction a standard
seniority pairing force has been used. The pairing strengths (for
protons and neutrons) used here were those of Ref.~\cite{Dude80}. To
avoid the well-known problems associated with the BCS treatment of
pairing in the region of low level densities, approximate number
projection was performed by means of the Lipkin-Nogami (LN) method 
\cite{Lipk76,Prad73,Naza90}. The energies of the local minima in the
potential energy surface (PES) were obtained by performing a
minimisation with respect to the $\beta_4$ and $\beta_6$ deformations
for each value of the quadrupole deformation $\beta_2$ 
      
Potential energy surface calculations (PES) for a series of isotopes in the 
Pb region (the Po, Pt, Pb and Hg isotopes) have been carried out
before \cite{Oros99,Beng89,Beng87,Satu91,Naza93,May77} 
with the aim of studying shape
coexistence in the Pb region. 
Properties of
the Pt, Hg, Pb, and Po nuclei are also discussed in the mass tables of
M\"oller {\it et al.}~\cite{Beng84,Moll81a,Moll88,Moll95,Moll97}.
A similar study has been carried out recently 
by R. Wyss \cite{Wyss-p}. 
This has been inspired, in particular, 
by the high-resolution mass measurements 
carried out by Schwarz {\it et al.} \cite{Schw01}.

\begin{figure}
\begin{center}
\mbox{\epsfig{file=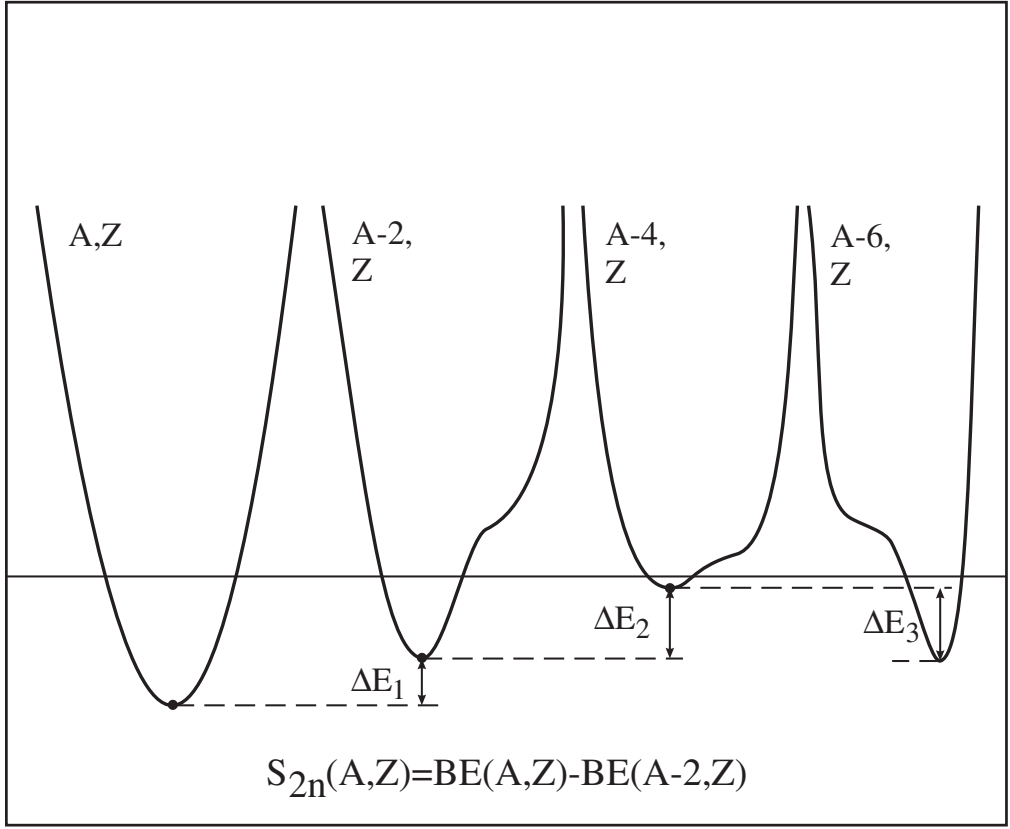,height=11.0cm,angle=0}}
\end{center}
\caption{Schematic representation of the method for calculating
  $S_{2n}$ in the study of PES using macroscopic-microscopic models.}
\label{fig-tpe-method}
\end{figure}

In the calculation of the PES, an absolute and several local minima
can occur, corresponding to a regular state 
($\varepsilon=0$) and to deformed states ($\varepsilon \not = 0$). 
Since here, we are interested in two-neutron separation energies, 
both the binding energies, 
{\it i.e.}~the absolute minimum, as well as the excitation energies  
of the local, deformed minima are important. In going from 
nucleus $(A,Z)$ to the adjacent nucleus, $(A-2,Z)$, in order to 
derive the $S_{2n}$ value, one has to take differences between 
the lowest total energy values for the nuclei that are considered 
(in absolute value). The method is illustrated in
Fig~\ref{fig-tpe-method}.  

We like to discuss in the next two paragraphs the precision with
which $S_{2n}$ can be calculated and this is important in the light
of later applications to nuclei in the Pb region.

When searching for minima in the potential energy surface (PES), 
as a function of deformation, a rather dense mesh
was used: $\Delta\beta_2 = 0.01$. The behavior obtained was smooth and
differences between energy values for neighbouring deformation
points were of the order of a few keV, at least close to the minimum.
Therefore, the computational error of the minimum energy can be
estimated as not exceeding 1 or 2 keV. Of course, the error introduced
by approximations of the method itself (e.g., by only approximate
fulfilment of the plateau condition or by use of simple Lipkin-Nogami
method for pairing correlations \cite{Lipk76,Prad73}) is much larger, 
closer to 1 MeV.
However, one can expect that taking only differences between energies
and not their absolute values cancels these errors substantially.

Inaccuracies could also be introduced through the dependence of
energy on Strutinsky's parameters $\gamma$ (``smoothing range'') and
$p$ (order of polynomial expansion used in smoothing procedure).
As discussed in detail in, e.g., Ref. \cite{Naza94}, even for not very
exotic nuclei, the plateau condition with respect to these parameters
is never fulfilled exactly. A change in $\gamma$ by
$\Delta\gamma\approx 0.2$ can induce a change in shell energy of
the order of 1 MeV in a rather irregular way depending on the detailed
structure of the single-particle spectrum. As a change in nucleon
number by 2 gives rise
to a change of shell energy of the order of up to a few MeV, the
uncertainties caused by lack of ideal plateau condition can be quite
substantial. Again, they are mainly smoothed out by taking
differences between adjacent even-even nuclei only. 
Such irregularities are most likely to show up
when deformation changes in passing from one nucleus to the heavier
or lighter by 2 nucleons, because the single particle-spectrum is then
different for the two nuclei. This latter effect might be  partly 
responsible for the presence of some irregularities 
(the smaller spikes as seen, e.g., in Fig. 20 for the Pt nuclei).

In the next subsection, we compare these results with 
the experimental values. One has to stress from the beginning
that PES calculations can, at best, give a 
qualitative description of binding energy differences ($S_{2n}$
values).
Dynamical effects obtained by solving a collective Schr\"{o}dinger 
equation are not taken care of (no mixing effects between close-lying  
$0^{+}$ are taken into account when making differences of PES 
values, purely), in the present discussion.

Going beyond the standard method of calculating PES, one needs to
account for mixing of the non-degenerate mean-field solutions. This
can be done using the generator coordinate method (GCM),
originally developed be Hill, Wheeler and Griffin \cite{Hill53,Grif57}.
Many-body states, obtained from HF+BCS or HFB calculations, are 
used to construct a more general ground state. 
At present, this can be done in a consistent way
using the same effective interaction. No systematic studies have been
performed as yet. Calculations starting from a solution in the
collective variables, $\beta$ and 
$\gamma$ and Euler angles, $\Omega$, have been carried out with
applications to the $N=20$ and $N=28$ shell closure by the group from
Bruy\`eres-le-Ch\^atel \cite{Peru00}. Likewise, calculations starting from
mean-field wave functions projected on angular momentum and particle
number using Skyrme interactions, have been carried out by the 
Orsay-Saclay-Brussels groups \cite{Valo00} . These calculations confirm
specific extra binding energy contributions like the ones discussed in 
section \ref{sec-ibm-mixing}. So, GCM calculations are becoming within
reach for more extended studies in {\it e.g.}~the Pb region because such
an approach accounts for the dynamics of the nuclear many-body problem. 
The shell-model, or,
symmetry truncated IBM studies are formulated in the laboratory frame,
conserving particle number and angular momentum from the beginning. In this
sense, the two approaches, though starting from a different zero-order
picture, essentially cover the same physics.

\subsection{Applications to the Pb region}
\label{sec-gen-pb}
The Pb region has shown a number of most interesting features when
moving into the neutron-deficient mass region. In the Pb nuclei,
low-lying $0^{+}$ excited states have been observed \cite{Cost00}. 
In the Po
nuclei, at the lowest mass numbers reached at present, clear
indications exist for a very low-lying $0^{+}$ state that even
might become the ground-state \cite{Oros99,Cost99}. 
In the Hg and Pt nuclei, clear-cut
evidence has accumulated for the presence of shape coexistence
\cite{Wood92}.  
In the Hg nuclei, the oblate shape is observed as being lowest in energy, 
even for the most neutron-deficient nuclei, whereas for the Pt nuclei,
a change from oblate into prolate shapes sets in around mass $A = 188$ and
the reverse path back from prolate into oblate around mass $A =
178$. Similar results have been derived before and discussed
\cite{Beng87}. 

It is the aim of the present study to explore how well
the mass dependence of the lowest energy minimum in a series of isotopes
correlates with the variations in the binding energy (using the 
two-neutron separation energies $S_{2n}$ as indicator) resulting from
the recent high-resolution mass measurements.

\begin{figure}
\begin{center}
\mbox{\epsfig{file=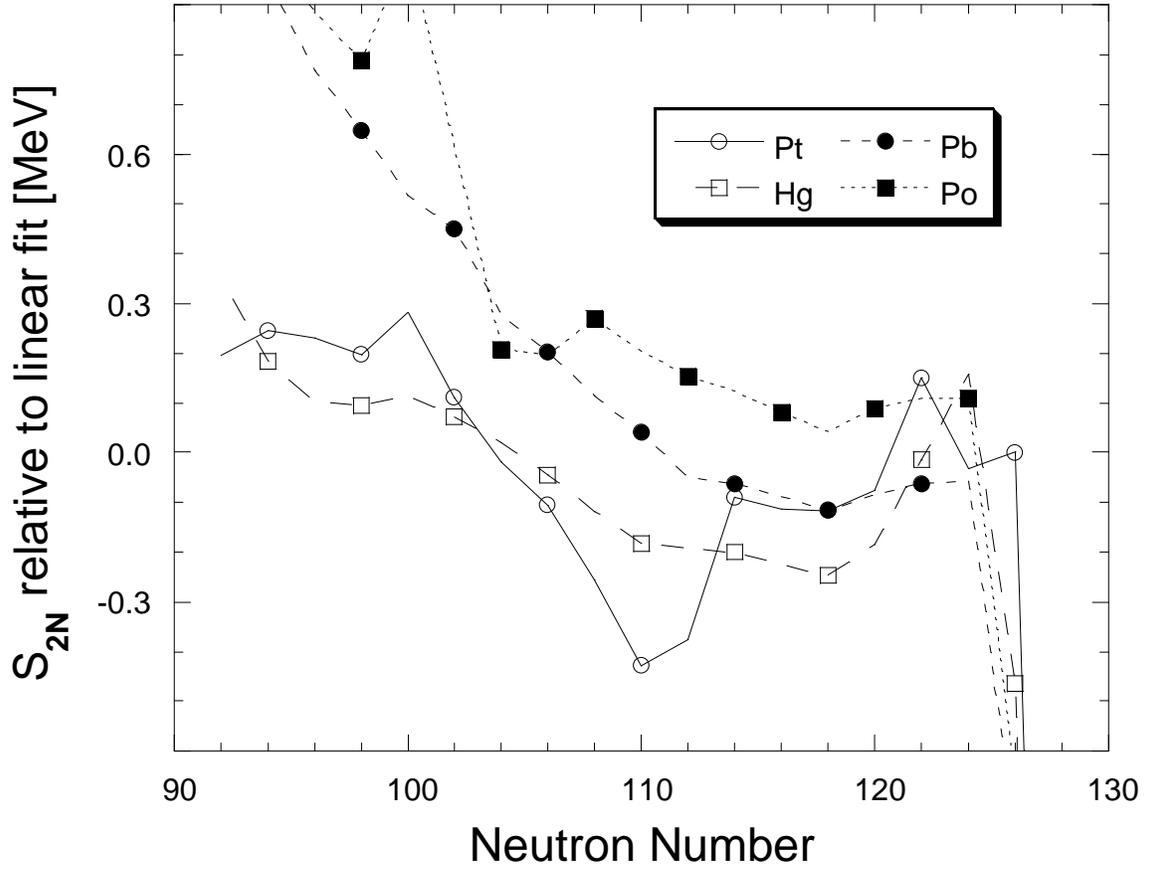,height=15.0cm,angle=-90}}
\end{center}
\caption{Theoretical $S_{2n}'$ values, $S_{2n}'(th)$, for Pt, Hg, Pb
  and Po using macroscopic-microscopic PES calculations.}
\label{fig-s2np-tpe-over}
\end{figure}

On the scale of a $S_{2n}$ plot (MeV energy scale) 
(see Fig.~\ref{fig-s2n-exp-isol}), no details can be seen of 
the ``intruder'' correlations (expected to be of the order of a few 
hundreds of keV). Therefore, we split up a $S_{2n}$ curve in two parts: 
a linear part and a part that contains local correlations (deformation
effects, specific mixing of configurations 
with the ground state, \ldots).
In order to visualise deviations from a linear behavior of the $S_{2n}$
values around neutron number, $100 \leq N \leq 110$ (the mid-shell region),
where nuclear shape coexistence and shape mixing is known to occur, the
linear curve was fitted to available experimental data outside of this
range (see \cite{Schw98,Kohl99,Schw01}). 
We shall concentrate on
these differences $S_{2n}'(exp) \equiv S_{2n}(exp)-S_{2n}(lin-fit)$ 
(and similarly in defining $S_{2n}'(th)$ as the difference of 
the theoretical value with the linear fit). In all following 
figures (unless stated  explicitly), we use these reduced
quantities. In Fig.~\ref{fig-s2np-tpe-over}, we give an overview of
the $S_{2n}'(th)$ values, obtained for the Pt, Hg, Pb and Po
nuclei. We shall give a more extensive discussion of these results in
comparing them with the data in the various subsections. Note the
differences with IBM in obtaining $S_{2n}(lin-fit)$ ($S_{2n}^{gl}$).

\subsubsection{The Pb $(Z=82)$ isotopes}
\label{sec-pb}
\begin{figure}
\begin{center}
\mbox{\epsfig{file=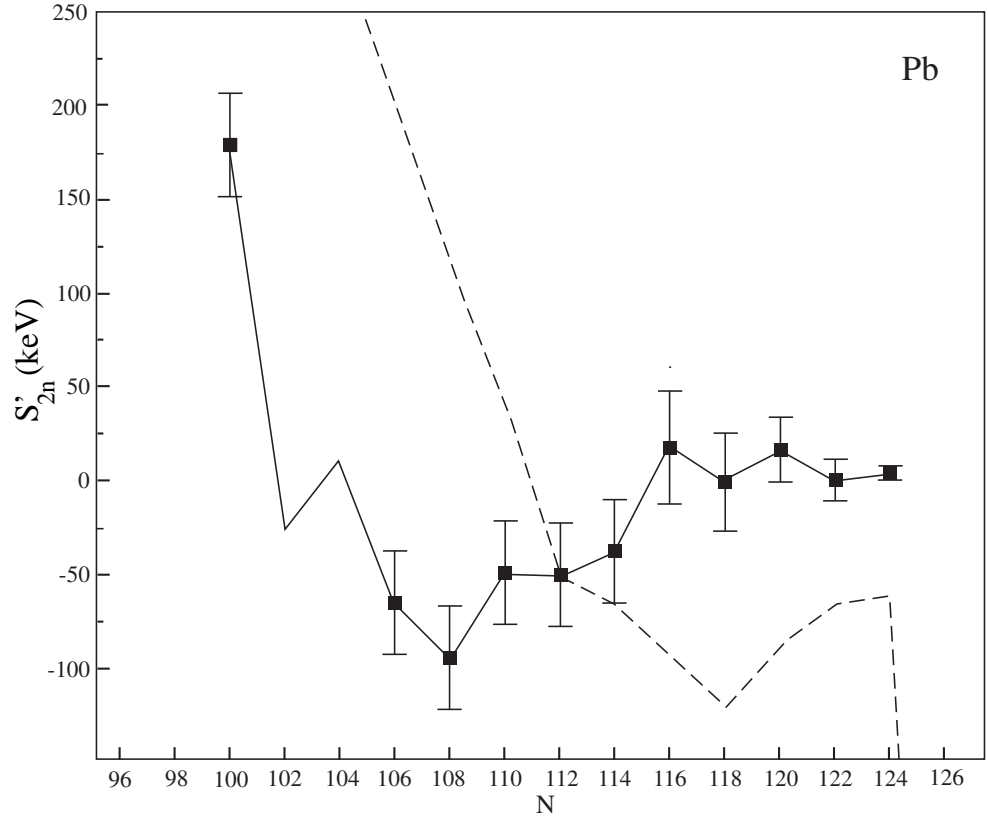,height=11.0cm,angle=0}}
\end{center}
\caption{$S_{2n}'$ for Pb. Comparison between experimental data (full
  line connecting dots) and PES results (dashed line).The two points
  (N=102,104) without data point are derived, containing at least one
  mass value obtained from mass systematics [45].
%  \cite{Audi97}.
  }
\label{fig-s2np-pb}
\end{figure}

The $S_{2n}'(exp)$ values are given in Fig.~\ref{fig-s2np-pb}. 
It is clear that down to the value at $N \simeq 110$, only a 
moderate lowering is observed (down to $\simeq$ $50$ keV). 
Beyond mid-shell neutron number ($N<104$), a rather
important increase in $S_{2n}'(exp)$ results. The relative variation 
in this quantity, moving out of the closed 
shell at $N=126$ towards mid-shell
and beyond, relative to the linear fit (which is approximating 
the local liquid-drop variation very well) is a reflection of the 
neutron shell-plus-pairing energy correction 
$\delta E=E_{shell}+E_{pair}$. 
These latter energy corrections 
causes the  neutron closed shell at $N=126$ 
to become more strongly bound (compared to a linear variation) and the 
mid-shell region to become less strongly bound (compared to the linear
variation). The fact that for the Pb nuclei, one has, at the same time, a
closed proton shell at $Z=82$, makes these variations relatively small on
an absolute scale and effects of deformation 
(occurrence of oblate and/or prolate shapes) cannot easily be 
observed on the present energy scale used.
The theoretical values $S_{2n}'(th)$, as plotted 
on the same figure \ref{fig-s2np-pb},
are derived starting from a deformed Woods-Saxon in calculating the 
PES curves. The theoretical curve takes a 
large jump down approaching the
neutron closed shell at $N=126$ and then remains moderately 
flat down to $N=114$.
Then, a smooth but steady increase is observed in passing through the 
mid-shell region (reflection of the behavior of $\delta E$). The
theoretical increase, though, advances the experimental increase by a 
couple of mass units. It is clearly very interesting that masses for even
lighter Pb isotopes could be determined to test this behavior.

\subsubsection{The Hg $(Z=80)$ and Pt $(Z=78)$ nuclei}
\label{sec-hg-pt}
\begin{figure}
\begin{center}
\mbox{\epsfig{file=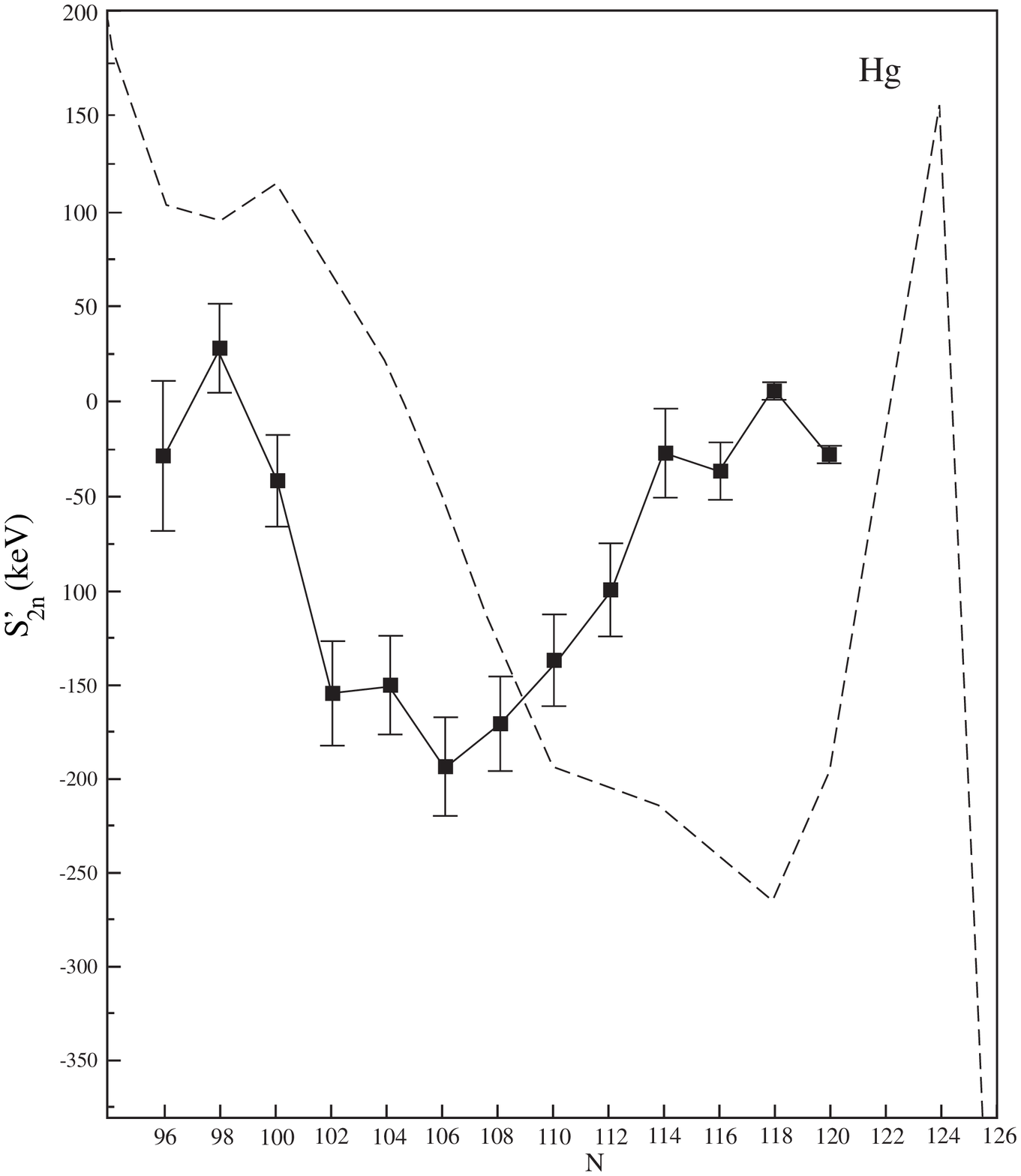,height=15.0cm,angle=0}}
\end{center}
\caption{$S_{2n}'$ for Hg. Comparison between experimental data (full
  line connecting dots) and PES results (dashed line).}
\label{fig-s2np-hg}
\end{figure}

The experimental reduced $S_{2n}$ plots, $S_{2n}'$, look rather
similar   
(see figures \ref{fig-s2np-hg} for Hg and \ref{fig-s2np-pt} for Pt
isotopes).  
Both show a systematic decrease in the separation energy for 
$N$ decreasing 
towards mid-shell at $N=104$. The plot for Hg shows a smooth valley with 
a minimum around $-200$ keV. Pt however, shows a sudden and steep
minimum  
of $-300$ keV for $^{186}_{\phantom{1}78}$Pt$_{108}$. 
Starting from mid-shell, the 
separation energies increase again for decreasing $N$.
 
One can check (see \cite{Wood92} for the specific energy spectra) 
that around 
mid-shell $N=104$ two band 
structures are present. Apart from the first one, 
known from heavy nuclei, 
that correspond to an oblate but weakly deformed structure, there appears 
another band that can be interpreted as a rotational 
spectrum corresponding 
with a prolate shape of larger deformation. In the Hg nuclei, this second 
band only approaches the ground state to $\simeq 400$ keV, whereas in the 
Pt nuclei, the prolate structure becomes
the ground-state band. As a conclusion, the even-even 
$^{178-186}$Pt$_{100-108}$ nuclei will have deformed ground states 
\cite{Schw01}. 

\begin{figure}
\begin{center}
\mbox{\epsfig{file=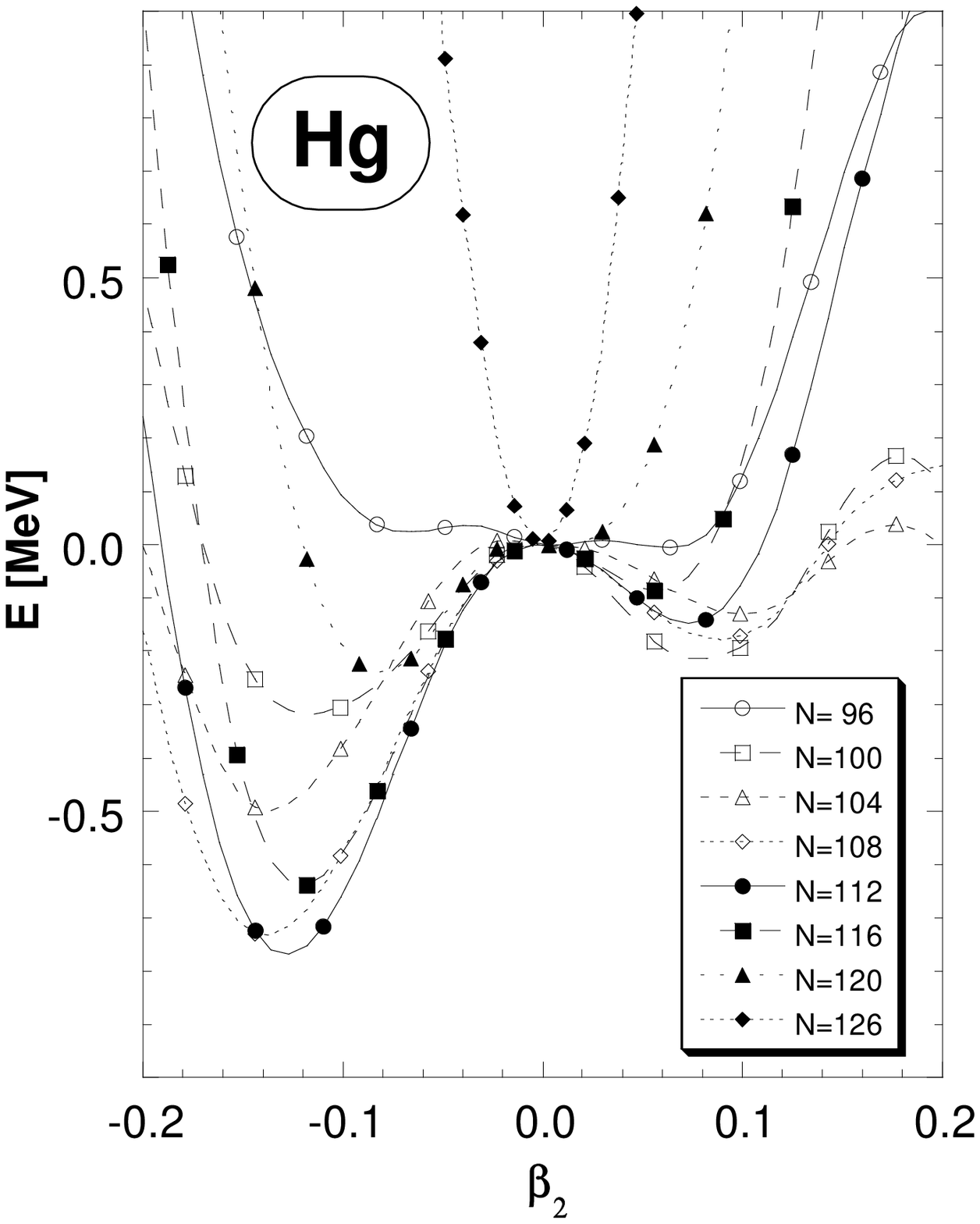,height=16.0cm,angle=0}}
\end{center}
\caption{Energy surface (PES) as a function of 
  the deformation parameter
  $\beta_2$ for different isotopes of Hg.}
\label{fig-s2np-hg-bis}
\end{figure}

\begin{figure}
\begin{center}
\mbox{\epsfig{file=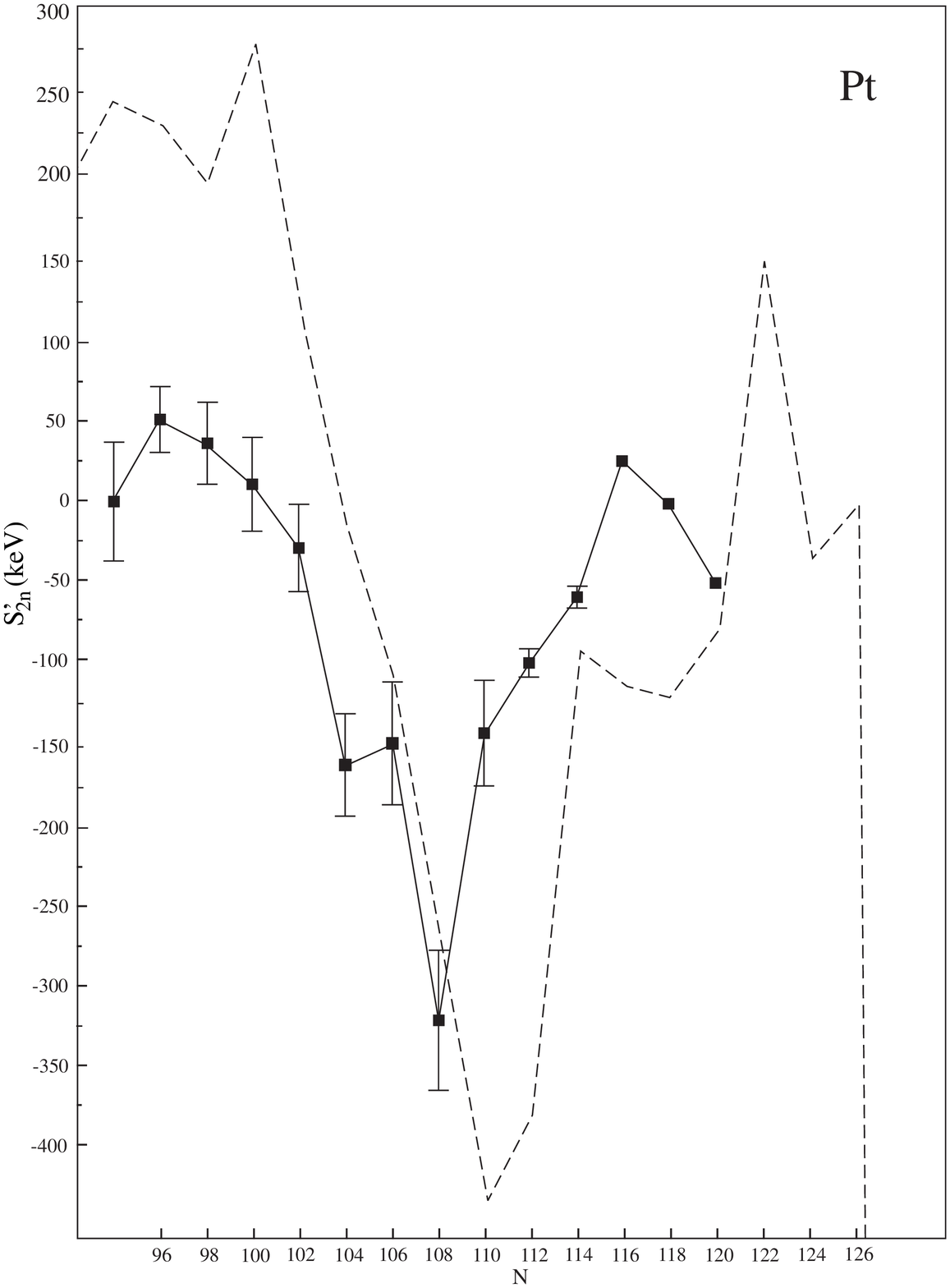,height=16.0cm,angle=0}}
\end{center}
\caption{$S_{2n}'$ for Pt. Comparison between experimental data (full
  line connecting dots) and PES results (dashed line).}
\label{fig-s2np-pt}
\end{figure}

For the Hg nuclei, we show in Fig.~\ref{fig-s2np-hg} 
a comparison between the 
reduced PES calculations, resulting in the 
theoretical $S_{2n}'(th)$
values and the corresponding experimental $S_{2n}'(exp)$ values. 
We observe 
a rather good overall agreement, except for the fact 
that in the theoretical
curve a more pronounced flat region is obtained for neutron numbers in the
interval $ 110 \leq N \leq 120 $, and the fact that below mid-shell, the
theoretical values become slightly positive. The first region 
($ 110 \leq N \leq 120 $) can be understood by the fact that the oblate
minimum for these Hg nuclei starts to develop, giving rise to a relative
increase in the binding energy over a spherical liquid-drop behavior (the
linear reference line at zero). The oblate minimum deepens down to
$N=112-114$ and then moves out again in the region $N=98-100$, 
approaching the liquid-drop reference line (see
Fig.~\ref{fig-s2np-hg-bis}).   

\begin{figure}
\begin{center}
\mbox{\epsfig{file=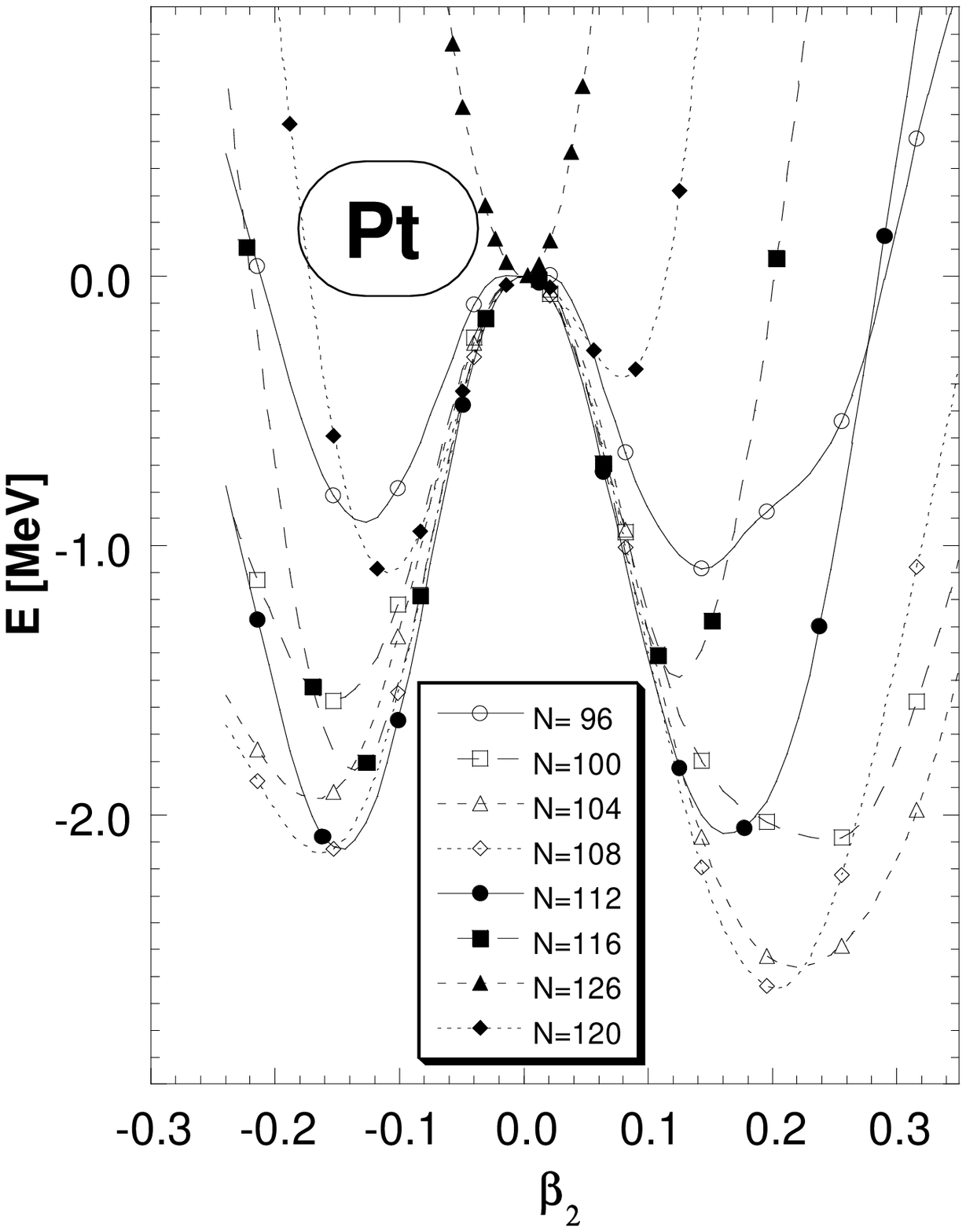,height=16.0cm,angle=0}}
\end{center}
\caption{Energy surface (PES) 
  as a function of the deformation parameter
  $\beta_2$ for different isotopes of Pt.}
\label{fig-s2np-pt-bis}
\end{figure}

For the Pt nuclei, the experimental and 
theoretical $S_{2n}'$ values are
plotted in Fig.~\ref{fig-s2np-pt}.  Concentrating on 
the theoretical curve, one can
relate the general structure with the PES curves and their variation 
with decreasing neutron number. At first, around 
neutron number $N=122-124$, the oblate minimum starts to 
develop, deepening
in going from $N=124$ down to $N=118$. At the same time, a 
prolate minimum takes shape and the minima for 
the oblate and prolate shape are almost
degenerate at $N=110$. The prolate minimum takes over and this minimum
starts becoming less deep at neutron 
number $N=106$. At $N \simeq  96$, the
prolate and oblate minima become degenerate again, but now as much less
pronounced and deep minima (see Fig.~\ref{fig-s2np-pt-bis}). 
This changing structure reflects the variation
of the theoretical $S_{2n}'$ values, a structure that is also observed
quite clearly in the experimental data.

\subsubsection{The Po $(Z=84)$ isotopes}
\label{sec-po}
\begin{figure}
\begin{center}
\mbox{\epsfig{file=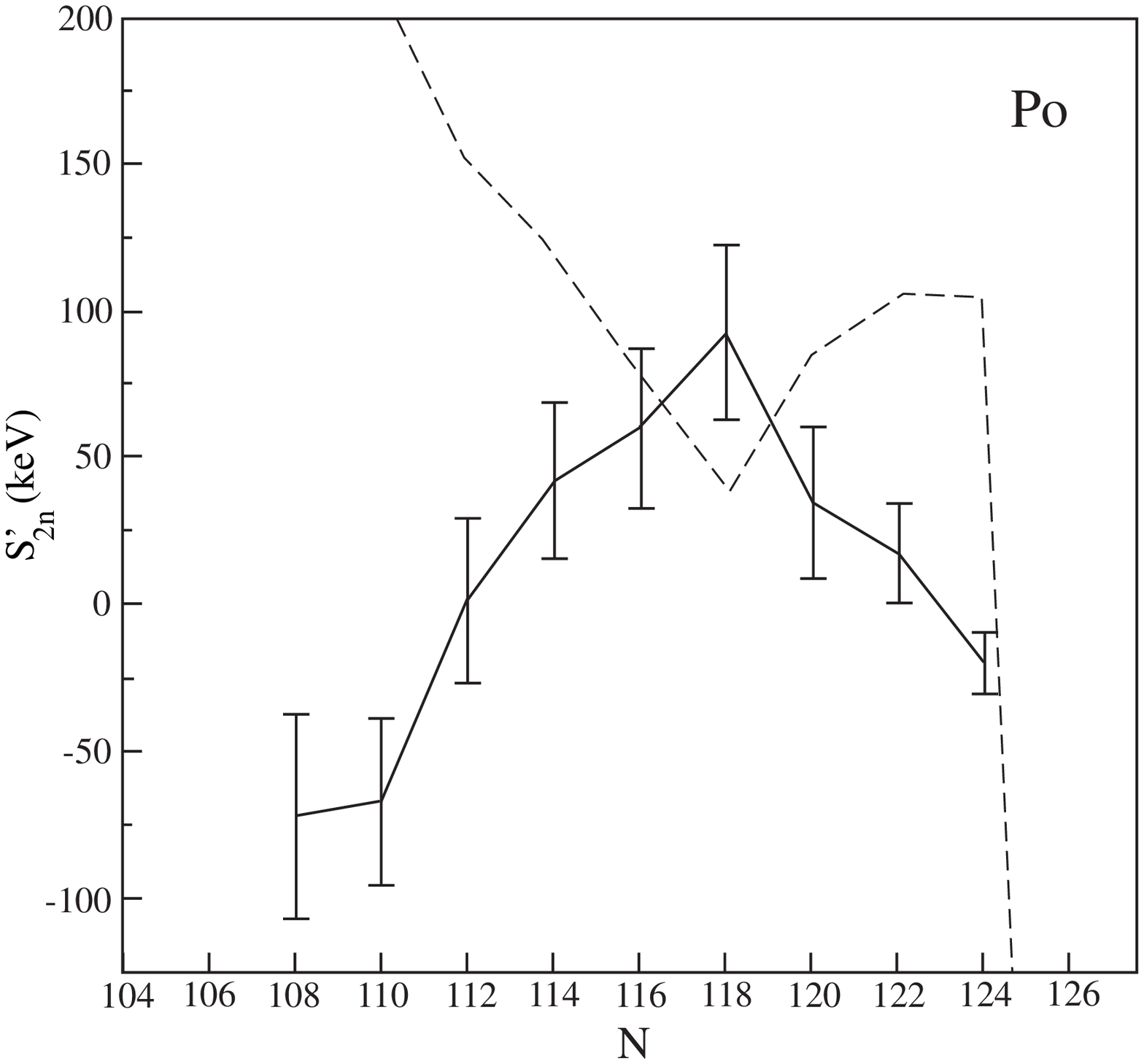,height=11.0cm,angle=0}}
\end{center}
\caption{$S_{2n}'$ for Po. Comparison between experimental data (full
  line connecting dots) and PES results (dashed line).}
\label{fig-s2np-po}
\end{figure}

The Po nuclei have recently been described using 
particle-core coupling and IBM studies \cite{Oros99,Cost99} and there,  
it has been discussed extensively that an intruder $0^{+}$ excited 
state is clearly dropping in energy with decreasing neutron number 
$N$ (from $N=118$ down to $N=108$) and might even
become the ground-state itself for these very neutron-deficient Po nuclei.
Inspecting now the experimental $S_{2n}'(exp)$ values 
(see Fig.~\ref{fig-s2np-po}),
it looks like the on-setting drop (below the reference linear fit line)
from $N=118$ down to $N=108$ is strongly correlated  with the 
above drop in energy of the intruder $0^{+}$ state.

\begin{figure}
\begin{center}
\mbox{\epsfig{file=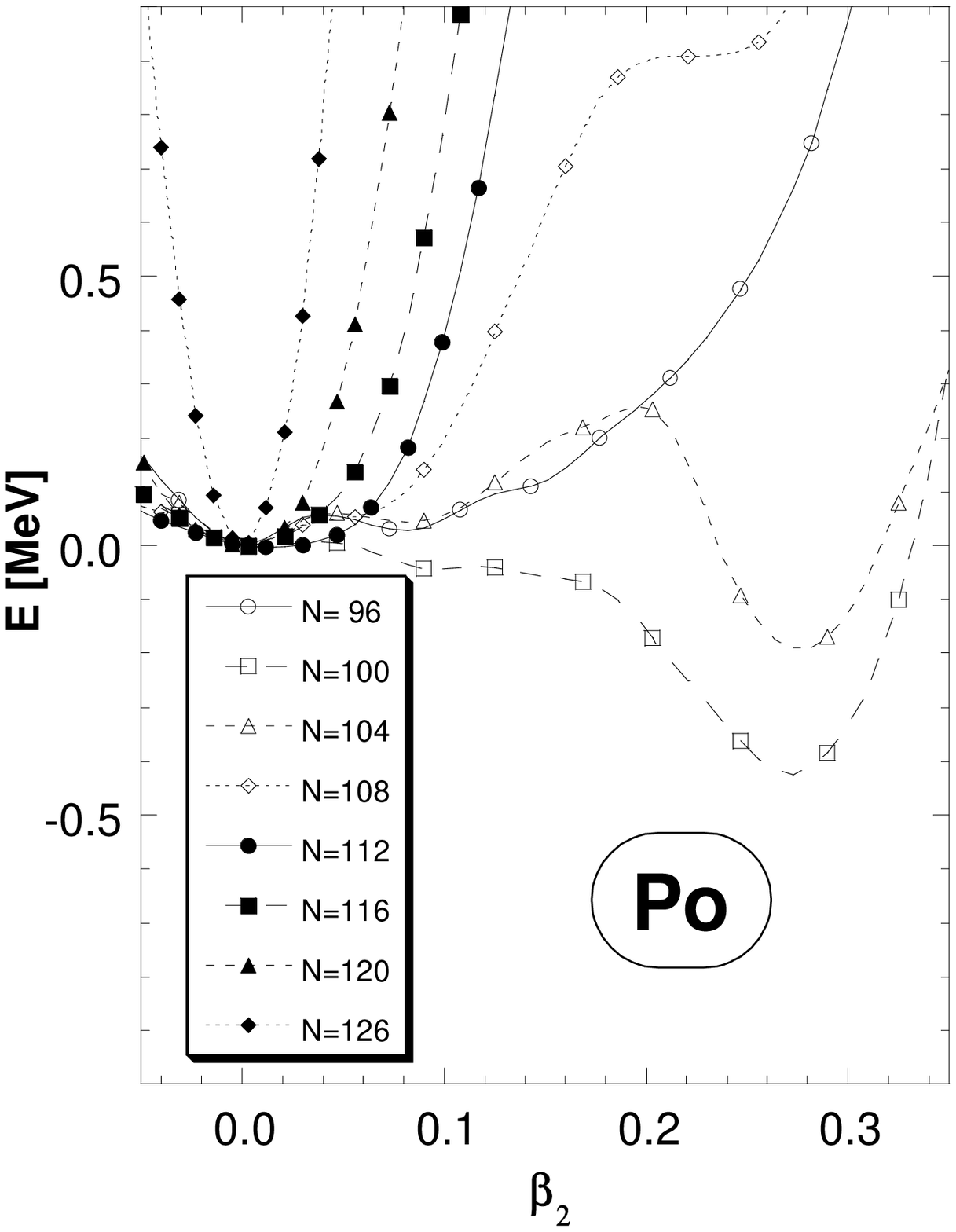,height=16.0cm,angle=0}}
\end{center}
\caption{Energy surface (PES) 
  as a function of the deformation parameter
  $\beta_2$ for different isotopes of Po.}
\label{fig-s2np-po-bis}
\end{figure}

In the results of the PES calculations (see also \cite{Oros99,Wyss-p} 
and Fig.~\ref{fig-s2np-po-bis}), the spherical minimum 
stays particularly stable
down to $N=118$ but then an oblate minimum starts coming 
on and deepens systematically, relative to the spherical minimum. 
In this respect, the comparison between the experimental and 
theoretical $S_{2n}'$ values goes rather well, down to $N=118$,  
in view of the energy scale used  on the Fig.~\ref{fig-s2np-po}. At 
neutron number $N=118$, the theoretical and experimental 
curves are going opposite ways. For the theoretical results, 
below $N=118$ down to $N=108$, the oblate minimum is developing 
relative to the spherical minimum with a prolate minimum quickly 
entering the picture that already is the lowest one 
(compared to the oblate minimum and the spherical point), see
Fig.~\ref{fig-s2np-po-bis}. 
The difference with the Hg and Pt nuclei is, that
in those cases (below $Z=82$), very quickly, the oblate and prolate
minima appear much below the energy corresponding with the spherical
point. In Po, on the contrary, this is not the case. 
Potential minima develop at
oblate and prolate shapes, but the spherical point dominates down to
$N=106$ (see Fig.~\ref{fig-s2np-po-bis}). 
It is this difference that causes the 
theoretical $S_{2n}'(theo)$ curve to move up 
(relative to the linear fit). The clear differences
then point out either, that the PES situation in the Po nuclei is not
so well reproduced, or, that dynamical effects, originating from 
mixing between the wave functions, localised at the 
various collective minima and the spherical point, play an important
role. The latter effect,
which is absent in any of the comparisons made in this section (the
Pb, Hg, Pt and Po nuclei) implies that in the present 
comparison we cannot
expect detailed agreement: only a qualitative correlation between 
trends in the theoretical and experimental $S_{2n}'$ can be expected.
Of course, the measurement of masses in the even more neutron-deficient
nuclei is extremely important. One might get access to the study of 
effects of (i) collective
(deformation versus spherical shape) correlations and (ii) specific
local configuration mixing inducing extra binding energy in the nuclear
ground-state configurations.

\subsection {A short conclusion} 

As a conclusion to section \ref{sec-intru}, 
we can say that mixing between 
intruder and regular states has an influence on both nuclear energy 
spectra and nuclear binding energies. 
However, we are still far from a detailed description of the consequences 
of mixing on the separation energies, as has become clear from 
the previous 
plots (see figures \ref{fig-s2np-pb},\ref{fig-s2np-hg},
\ref{fig-s2np-pt}, and \ref{fig-s2np-po}).

Even though the many parameters appearing in the
macroscopic-microscopic model to evaluate nuclear masses and PES,
those parameters are fitted to an enormous mount of data in a complex
fitting procedure; they remain untouched after that. Only a few of
them (see also section \ref{sec-ldm-vall}) are really fitted to all
masses; the other ones are fitted to selected data sets. Therefore, the
masses derived this way (and derived $S_{2n}$ values) cannot be used
to account for all localised nuclear structure correlations (at and
near closed shells, the precise onset of regions of deformation), which
is also not the aim of the many mass studies
\cite{Beng84,Moll81a,Moll88,Moll95,Moll97,Abou95,Lala99,Gori01}. 
Only after including dynamical effects (like {\it e.g.}~the GCM approach)
can one expect to cover both, the full global and local mass behavior. 

It should be stressed that in the case of the IBM calculations, the 
parameters of the local Hamiltonian 
have been chosen independently from the ``$S_{2n}$ problem''. 
They were obtained from an independent fit of the energy spectra in this 
mass region for both, the regular and intruder states. 
As far as the two-neutron 
separation energies are concerned, the IBM results can count as a 
prediction for the actual $S_{2n}'$ values.

\section{Conclusions}
\label{sec-conclu}
In the present paper, in which we have studied nuclear binding energies 
and their global properties over a large region of the nuclear mass
table, we also concentrated on local deviations from a smooth behavior
and made use of the two-neutron separation energy, $S_{2n}$, as an
important property to explore the nuclear mass surface. The latter
local variations could stem from the presence of shell or sub-shell
closure, the appearance of a localised region of deformation or might
originate in specific configuration mixing with the ground state that
causes local increased binding energies to show up.

The very recent high-resolution measurements that have been carried out,
in particular at the ISOLTRAP and MISTRAL set-ups at ISOLDE/CERN have
allowed to study nuclear masses with an unprecedented  
precision of $10^{-5}$ and as
such brings the interest of mass measurements from tests of global
mass formulae or HF(B) studies into a realm that allows tests of
shell-model calculations.

Here, we have discussed up to what level calculations -  making
use of global macroscopic models, macroscopic-microscopic
calculations, the shell-model and the Interacting 
Boson Model -  can give a correct overall description of the nuclear mass
surface (along the region of the valley of stability as well as for
series of isotopes). It has become clear that, if one starts from a
simple liquid-drop approach, the observed almost linear drop
in the $S_{2n}$ value is accounted essentially through the asymmetry
term. This term causes nuclei to become less bound when moving out of
the region where $Z \approx \frac{A}{2}$ in a systematic way and even
turns to a linear variation in (\ref{ldm-a}) 
when the neutron excess is becoming
really large. The liquid drop approach is able to give the correct overall
mass dependence in $S_{2n}$ along the stability line as well as for long
series of isotopes. It is observed though that the experimental slope is
somewhat less pronounced compared to the liquid drop behavior (see 
Fig.~\ref{fig-ldm-contrib}), but here, more sophisticated
macroscopic-microscopic calculations have resulted into impressive
results. 

The above features also result from a 
shell-model approach in which
we treat a given mass region approximately starting from a reference
(doubly)-closed shell nucleus and have the valence nucleons filling 
a single-$j$ shell model orbital. Using a zero-range 
$\delta$ interaction, a linear
variation with the number of nucleons, $n$, in 
describing the binding energy
$BE(j,n)$ results. For more general interactions, still keeping seniority,
$v$, as a good quantum number, a linear plus quadratic $n$ dependence is
obtained with the coefficient of the quadratic part contributing with 
a repulsive
component to the total binding energy of the shell. This term is similar
in nature and relative magnitude to the asymmetry term of the liquid-drop
model description. Finally, a linear drop in the value 
of $S_{2n}$ results.

There is a clear need to do better and try more detailed shell-model
calculations. At present, the limitations of standard 
large-scale diagonalization
constrain the calculations to the $fp$ shell. 
Recent, new developments, starting from diagonalizations in a basis
generated from Monte-Carlo sampling of the essential nuclear degrees
of freedom (Monte-Carlo shell model diagonalization: MCSMD) have
resulted in highly encouraging results that may open new possibilities
to cover both, global and local nuclear properties in a consistent and
unified way \cite{Honn95,Shim01}).  
  
There are remaining problems connected with deriving
absolute binding energies since one needs a good 
description for the variation
with $A$ or $N$ of the single-particle energies 
$\epsilon_{j}$. Some prescriptions are in use \cite{Dufo96,Dufl99},
but there remains a difficulty for pure shell-model studies.

In plotting the value of $S_{2n}$ versus the number of nucleon pairs,
the shape is close to linear. Deviations, however,  
show up that must be due to the
subsequent filling of a number of single-particle orbitals and to the 
correlation energy that results from the interactions in which pairing and
proton-neutron forces play a major role.  A pair approximation, as used
within the Interacting Boson Model, can then be used to take into account
both the global and local components of 
nuclear binding energy. We have
discussed a procedure which starts from simultaneous 
treating binding energies
and excitation energies in extracting parameters for the 
linear and quadratic
$U(6)$ Casimir invariant operators. This approach bases on the
assumption that the added global part to the IBM result is the same
for a chain of isotopes. It should be stressed that the linear part will
change when changing between major shells and even when crossing the
mid-shell region. This latter fact turns out to be an 
intrinsic deficiency of the
IBM due to the fact that the Pauli principle is included only in an 
approximate way (there is no reference any more to the 
subsequent filling of
a set of single-particle orbitals with a maximal number of nucleons). 
In section \ref{sec-calc-ab}, we have reported a detailed
prescription in order to obtain a consistent 
description of binding energies, 
energy spectra and transition rates in the framework of the IBM. 

In a second part of the paper, we have concentrated particularly on
local deviations from the above global description. The possibility to
study nuclear masses with the highest possible 
precision has become available
over the last years, in particular at the ISOLTRAP and MISTRAL set-ups at
ISOLDE/CERN. Here, precisions of 
the order of $30$ keV on a total mass of a
heavy Pb nuclei ($\approx$ $1600$ MeV) is reached. 
This has given rise to
a number of unexpected features in the masses of neutron-deficient nuclei
in the region of Pt, Hg, Pb, Po, Rn, Ra \cite{Schw98,Kohl99,Schw01}. 
Before, similar local deviations had been observed in the region 
of light $N=20$ nuclei for Na, Mg 
\cite{Warb90,Wood92,Utsu99,Azai00,Caur98,Rodr00}. 

In the present paper, we have pointed out the necessity to 
incorporate configuration mixing of the regular 
ground state with low-lying
$0^{+}$ intruder states that approach the $0^{+}$ ground-state in the
neutron mid-shell region ($N \approx 104$) for nuclei near closed-shell
configurations. We have carried out detailed 
calculations for the Po nuclei.
Similar calculations for the whole Pb region will be carried out
elsewhere in a consistent
way. At the same time, calculations for the potential energy surfaces(PES)
of the given nuclei in the Pb region have been performed, using
the macroscopic-microscopic model with the universal deformed 
Woods-Saxon parametrisation. Here, the various
shapes: spherical configuration, oblate and prolate deformed shapes and
their relative ordering, as a function of neutron number, is instrumental
in understanding local ground-state energy deviations from the background
liquid-drop behavior. We observe a good correlation between the
experimental values of $S_{2n}'$ and the calculated ones 
on the $100$ keV-scale. These
results are encouraging in the light of lack of dynamical effects:
we just compare energy minima for different shapes without taking into
account mixing that will inevitably occur between such close-lying states.

Resuming, we have shown, in a first part, that both, a liquid-drop
approach (macroscopic-microscopic models in general) as well as 
the shell model and the IBM describe the global part
of the $S_{2n}$ value essentially identical. The linear drop is mainly
connected to the asymmetry term (in the LDM), 
the quadratic terms (in the shell model)
and the quadratic $U(6)$ Casimir invariant 
(in the IBM), but all three contain the same physics.
Both, the overall drop in $S_{2n}$ for the whole mass region, as well as
in specific long isotopic series are well accounted for. 
The experimental
drop is overall less steep. In particular, in the case of the IBM we
have shown that it is possible to give a consistent description of
ground-state and excited-state properties. This description is able
to reproduce the experimental $S_{2n}$ values rather well.
Finally, in a third part, deviations in nuclear binding from
the global trend are showing up in various localised regions. In the Pb
region, it is most probably 
the effect of mixing of low-lying intruder configurations
(oblate and/or prolate shape configurations) into the ground-state that
turns out to be responsible for increased binding energies in the
neutron-deficient region. Using configuration mixing in the IBM, detailed
studies can be carried out and a consistent study is planned for the
Pb region and for other (sub)shell-closures. The PES study of static
properties on the other hand is able to give a good 
guidance to the interpretation
of the specific deviations that have been observed in the Pb region.

Thus nuclear mass measurements are becoming increasingly important since 
they have
progressed now to the level of testing microscopic studies (shell-model
effects, localised zones of nuclear deformation, \ldots). 
This will become clearly an important line of research in future
projects.   

\section{Acknowledgements}

The authors like to thank Stefan Schwarz, Alexander Kohl 
and George Bollen for
discussions at ISOLDE and to the ISOLDE group for 
much discussions during the
early phase of this work. We also thank R. Wyss for communicating PES
calculations in the Pb region. We are grateful 
to R.F.~Casten, F.~Iachello, W.~Nazarewicz, P.~Van Isacker, and
J.L.~Wood for much input, inspiration,
and critical discussions. Finally, we thank the ``FWO-Vlaanderen'',
NATO for the research grant CRG96-0981. One of us (T. W.) 
has benefitted from
the Flemish-Polish bilateral projects BIL 97/174B15.98 
and BIL 01/174B15.01. 
K. H. likes to thank J. \"{A}yst\"{o}
and the ISOLDE collaboration at CERN for an interesting 
stay and support during the final stages  of this work.
Finally, they are grateful to the referees for much constructive
criticism.

\newpage

\begin{table}
\caption{Parameters for an IBM Hamiltonian for Xe isotopes. 
With $\epsilon_d$ in keV, $\kappa=30$ keV and $\kappa'=0$.}
\begin{center}
\begin{tabular}{ccc|ccc}
$A$ &$N_\nu$&$\epsilon_d$   &$A$ &$N_\nu$&$\epsilon_d$\\  
\hline
114 &7      &67             &130 &5      &62          \\  
116 &8      &67             &132 &4      &70          \\ 
118 &9      &66             &134 &3      &80          \\   
120 &10     &70             &138 &3      &61          \\   
122 &9      &65             &140 &4      &45          \\   
124 &8      &62             &142 &5      &42          \\   
126 &7      &60             &144 &6      &45          \\   
128 &6      &60             &~   &~      &~           \\   
\end{tabular}
\end{center}
\label{tab-xe-ham}
\end{table}

\begin{thebibliography}{99}
\bibitem{Ring80}
P.~Ring and P.~Schuck, {\em The nuclear many-body problem}
(Springer-Verlag, New York, Heidelberg, Berlin, 1980).

\bibitem{Nils95}
S.G.~Nilsson and I.~Ragnarson,
{\em Shapes and shells in the nuclear structure}.
(Cambridge University Press, Cambridge, 1995).

\bibitem{Waps58}
A.H.~Wapstra, {\em Handbuch der Physik} Vol.~1, (1958).

\bibitem{Waps71}
A.H.~Wapstra and N.B.~Gove, Nucl.\ Data Tables {\bf9}, 267 (1971). 

\bibitem{Myer67}
W.D.~Myers and W.J.~Swiatecki, Ark.\ Fys.\ {\bf36}, 343 (1967).

\bibitem{Myer69}
W.D.~Myers and W.J.~Swiatecki, Ann.\ Phys.\ (NY) {\bf55}, 395 (1969). 

\bibitem{Moll81}
P.~M\"oller and J.R.~Nix, Nucl.\ Phys.\ A {\bf361}, 117 (1981).

\bibitem{Beng84}
R.~Bengtsson, P.~M\"oller, J.R.~Nix, and J.-Y.~Zhang, Phys.\ Scr.\
{\bf29}, 402 (1984).

\bibitem{Moll81a}
P.~M\"oller and J.R.~Nix, At.\ Data Nucl.\ Data Tables {\bf26}, 165
(1981). 

\bibitem{Moll88}
P.~M\"oller and J.R.~Nix, At.\ Data Nucl.\ Data Tables {\bf39}, 213
(1988). 

\bibitem{Moll95}
P.~M\"oller, J.R.~Nix, W.D.~Myers, and W.J.~Swiatecki, 
At.\ Data Nucl.\ Data Tables {\bf59}, 185 (1995).

\bibitem{Moll97}
P.~M\"oller, J.R.~Nix, and K.-L.~Kratz,
At.\ Data Nucl.\ Data Tables {\bf66}, 131 (1997).

\bibitem{Abou95}
Y.~Aboussir, J.M.~Pearson, A.K.~Dutta, and F.~Tondeur,
At.\ Data Nucl.\ Data Tables {\bf61}, 127 (1995).

\bibitem{Lala99}
G.A.~Lalazissis, S.~Raman, and P.~Ring,
At.\ Data Nucl.\ Data Tables {\bf71}, 1 (1999).
 
\bibitem{Gori01}
S.~Goriely, F.~Tondeur, and J.M.~Pearson,
At.\ Data Nucl.\ Data Tables {\bf77}, 311 (2001).
 
\bibitem{Boll96}
G.~Bollen {\it et~al.},
Nucl.\ Instr.\ and Meth.\ A {\bf368}, 675 (1996).

\bibitem{Raim97}
H.~Raimbault-Hartmann {\it et~al.},
Nucl.\ Instr.\ and Meth.\ B {\bf126}, 378 (1997).

\bibitem{Schw98}
S.~Schwarz,
PhD.~Thesis, University of Mainz, (1998), unpublished.

\bibitem{Kohl99}
A.~Kohl, PhD.~Thesis, University of Heidelberg, (1999), unpublished.

\bibitem{Schw01}
S.~Schwarz {\it et~al.}, Nucl.\ Phys.\ A (2001), in press.

\bibitem{Weiz35}
C.F.~von Weizs\"acker,
Z.\ Phys.\ {\bf96}, 431 (1935).

\bibitem{Beth36}
H.A.~Bethe and R.F.~Bacher,
Rev.\ Mod.\ Phys.\ {\bf8}, 193 (1936).

\bibitem{Krup00}
A.T.~Kruppa, M.~Bender, W.~Nazarewicz, P.-G.~Reinhard, T.~Vertse, and 
\'Cwiok, Phys.\ Rev.\ C {\bf61}, 03413 (2000).

\bibitem{Shal63}
A.~de~Shalit and I.~Talmi,
{\em Nuclear shell theory}
(Academic Press, New York and London, 1963).

\bibitem{Talm93}
I.~Talmi,
{\em Simple models of complex nuclei}
(Harwood Academic Publishers, 1993).

\bibitem{Talm71}
I.~Talmi, Nucl.\ Phys.\ A {\bf172}, 1 (1971).

\bibitem{Wood-p}
J.L.~Wood, private communication and to be published.

\bibitem{Dufo96}
M.~Dufour and A.P.~Zuker, Phys.\ Rev.\ C {\bf54}, 1641 (1996).

\bibitem{Dufl99}
J.~Duflo and A.P.~Zuker, Phys.\ Rev.\ C {\bf59}, R2347 (1999).

\bibitem{Brow88}
B.A.~Brown and B.H.~Wildenthal, 
Ann.\ Rev.\ Nucl.\ Part.\ Sci {\bf38}, 29 (1988).

\bibitem{Warb90}
E.K.~Warburton, J.A.~Becker, and B.A.~Brown,
Phys.\ Rep.\ {\bf41}, 1147 (1990).

\bibitem{Wood92}
J.L.~Wood, K.~Heyde, W.~Nazarewicz, M.~Huyse, and P.~Van~Duppen,
Phys.\ Rep.\ {\bf215}, 101 (1992).

\bibitem{Iach87}
F.~Iachello and A.~Arima,
{\em The interacting boson model}
(Cambridge University Press, Cambridge, 1987).

\bibitem{Fran94}
A.~Frank and P.~Van~Isacker,
{\em Algebraic methods in molecular and nuclear structure physics}
(Wiley-Interscience, 1994).

\bibitem{Cast88}
R.F.~Casten and D.D.~Warner,
Rev.\ Mod.\ Phys.\ {\bf60}, 389 (1988).

\bibitem{Garc01a}
J.E.~Garc\'{\i}a-Ramos, C.~De~Coster, R.~Fossion, and K.~Heyde,
Nucl.\ Phys.\ A {\bf688}, 735 (2001).

\bibitem{Chou97}
W.-T.~Chou, N.V.~Zamfir, and R.F.~Casten,
Phys.\ Rev.\ C {\bf56}, 829 (1997).

\bibitem{Warn82}
D.D.~Warner and R.F.~Casten,
Phys.\ Rev.\ Lett.\ {\bf48}, 1385 (1982).

\bibitem{Stru67}
V.M.~Strutinsky, Nucl.\ Phys.\ A {\bf95}, 420 (1967).

\bibitem{Stru68}
V.M.~Strutinsky, Nucl.\ Phys.\ A {\bf122}, 1 (1968).

\bibitem{Brus77}
P.J.~Brussaard and P.W.M.~Glaudemans,
{\em Shell-model applications in nuclear spectroscopy}
(North-Holland, Amsterdam, 1977).

\bibitem{Heyd94}
K.~Heyde, {\em The nuclear shell-model} (Springer-Verlag, Berlin,
Heidelberg, New York, 1994).

\bibitem{Audi93}
G.~Audi and A.H.~Wapstra,
Nucl.\ Phys.\ A {\bf565}, 193 (1993).

\bibitem{Audi95}
G.~Audi and A.H.~Wapstra,
Nucl.\ Phys.\ A {\bf595}, 409 (1995).

\bibitem{Audi97}
G.~Audi, D.~Bersillon, J.~Blachot, and A.H.~Wapstra
Nucl.\ Phys.\ A {\bf624}, 1 (1997). Database
http://csn.www.in2p3.fr/amdc.

\bibitem{Beck00}
B.~Beck, {\it et~al.}, Eur.\ Phys.\ J.\ A {\bf8}, 307 (2000).
 
\bibitem{Heyd83}
K.~Heyde, P.~Van~Isacker, M.~Waroquier, J.L.~Wood, and R.A.~Meyer,
Phys.\ Rep.\ {\bf102}, 292 (1983).

\bibitem{Duva81}
P.D.~Duval and B.R.~Barrett, Phys.\ Lett.\ B {\bf100}, 223 (1981).

\bibitem{Duva82}
P.D.~Duval and B.R.~Barrett,
Nucl.\ Phys.\ A {\bf376}, 213 (1982).

\bibitem{Barf83}
A.F.~Barfield, B.R.~Barrett, K.A.~Sage, and P.D.~Duval,
Z.\ Phys.\ A {\bf311}, 205 (1983).

\bibitem{Hard97}
M.K.~Harder, K.T.~Tang, and P.~Van~Isacker,
Phys.\ Lett.\ B {\bf405}, 25 (1997).

\bibitem{Oros99}
A.~Oros, K.~Heyde, C.~De~Coster, B.~Decroix, R.~Wyss, B.R.~Barrett,
and P.~Navratil, Nucl.\ Phys.\ A {\bf645}, 107 (1999).

\bibitem{Cost99}
C.~De~Coster, K.~Heyde, B.~Decroix, J.L.~Wood, J.~Jolie, 
and H.~Lehmann, Nucl.\ Phys.\ A {\bf651}, 31 (1999).

\bibitem{Dele93a}
M.~D\'el\`eze, S.~Drissi, J.~Kern, P.A.~Tercier, J.P.~Vorlet, 
J.~Rikovska, T.~Otsuka, S.~Judge, and A.~Williams,
Nucl.\ Phys.\ A {\bf551}, 269 (1993).

\bibitem{Dele93b}
M.~D\'el\`eze, S.~Drissi, J.~Jolie, J.~Kern, and J.P.~Vorlet,
Nucl.\ Phys.\ A {\bf554}, 1 (1993).

\bibitem{Lehm97}
H.~Lehmann, J.~Jolie, C.~De~Coster, K.~Heyde, B.~Decroix, 
and J.L.~Wood, Nucl.\ Phys.\ A {\bf621}, 767 (1997).

\bibitem{Heyd87}
K.~Heyde, J.~Jolie, J.~Moreau, J.~Ryckebusch, M.~Waroquier, 
P.~Van~Duppen, M.~Huyse, and J.L.~Wood,
Nucl.\ Phys.\ A {\bf621}, 767 (1997).

\bibitem{Utsu99}
Y.~Utsuno, T.~Otsuka, T.~Mizusaki, and M.~Honma,
Phys.\ Rev. C {\bf60}, 054315 (1999).

\bibitem{Azai00}
F.~Azaiez, Phys.\ Scr.\ {\bf88}, 118 (2000).

\bibitem{Caur98}
E.~Caurier, F.~Nowacki, A.~Poves, and J.~Retamosa,
Phys.\ Rev. C {\bf58}, 2033 (1998).

\bibitem{Rodr00}
R.R.~Rodr\'{\i}guez-Guzm\'an, J.L.~Egido, and L.M.~Robledo,
Phys.\ Rev. C {\bf62}, 054319 (2000).

\bibitem{Cost00}
C.~De~Coster, B.~Decroix, and K.~Heyde,
Phys.\ Rev.\ C {\bf61}, 067306 (2000).

\bibitem{Kibe94}
T.~Kib\'edi, G.D.~Dracoulis, A.P.~Byrne, P.M.~Davidson, 
and S.~Kuyucak, Nucl.\ Phys.\ A {\bf567}, 183 (1994).

\bibitem{Dupp87}
P.~Van~Duppen, E.~Coenen, K.~Deneffe, M.~Huyse, and J.L.~Wood, 
Phys.\ Rev.\ C {\bf35}, 1861 (1987).

\bibitem{Dupp90}
P.~Van~Duppen, M.~Huyse, and J.L.~Wood, 
J.\ Phys.\ G {\bf16}, 441 (1990).

\bibitem{Brac72}
M.~Brack, J.~Damgaard, A.S.~Jensen, H.C.~Pauli, 
V.M.~Strutinsky, and Wong C.Y, Rev.\ Mod.\ Phys.\ {\bf44}, 320
(1972). 

\bibitem{Beng89}
R.~Bengtsson and W.~Nazarewicz, 
Z.\ Phys.\ A {\bf334}, 269 (1989).

\bibitem{Cwio87} S.~\'Cwiok, J.~Dudek, W.~Nazarewicz, J.~Skalski, 
T.R.~Werner, Comp. Phys. Comm. {\bf 46} 379 (1987).

\bibitem{Dude82} J.~Dudek, Z.~Szyma\'nski, T.R.~Werner, A.~Faessler, 
C.~Lima, Phys.\ Rev.\ C {\bf 26} 1712 (1982).

\bibitem{Dude81}
J.~Dudek, Z.~Szyma\'nski, T.R.~Werner,  Phys.\ Rev.\ C {\bf23} 920
(1981).

\bibitem{Beng87}
R.~Bengtsson, T.~Bengtsson, J.~Dudek, G.~Leander,  W.~Nazarewicz, and
J.-Y.~Zhang, Phys.\ Lett.\ B {\bf183}, 1 (1987).

\bibitem{Satu91}
W.~Satula, S.~\'Cwiok, W.~Nazarewicz, R.~Wyss and A.~Johnson, Nucl.\
Phys.\ A {\bf529}, 289 (1991).

\bibitem{Naza93}
W.~Nazarewicz, Phys.\ Lett.\ B {\bf305}, 195 (1993).

\bibitem{May77}
F.R.~May, V.V.~Pashkevich, and S.~Frauendorf,
Phys.\ Lett.\ B {\bf68}, 113 (1977).

\bibitem{Dude80} 
J.~Dudek, A.~Majhofer, J.~Skalski, J.\ Phys.\ G {\bf 6} 447 (1980).

\bibitem{Lipk76}
H.J.~Lipkin, Ann.\ Phys.\ (NY) {\bf31}, 525 (1976).

\bibitem{Prad73}
H.C.~Pradhan, Y.~Nogami, and J.~Law, Nucl.\ Phys.\ A {\bf201}, 357
(1973).

\bibitem{Naza90}
W.~Nazarewicz, M.A.~Riley, and J.D.~Garrett, Nucl.\ Phys.\ A 
{\bf512}, 61 (1990).

\bibitem{Wyss-p}
R.~Wyss, private communication.

\bibitem{Naza94}
W.~Nazarewicz, T.R.~Werner, J.~Dobaczewski, Phys.\ Rev.\ C 
{\bf50}, 2860 (1994).

\bibitem{Hill53}
D.L.~Hill and J.A.~Wheeler, Phys.\ Rev.\ {\bf89}, 1102 (1953).

\bibitem{Grif57}
J.J.~Griffin and J.A.~Wheeler, Phys.\ Rev.\ {\bf108}, 311 (1957).

\bibitem{Peru00}
S.~P\`eru, M.~Girod, and J.F.~Berger, Eur.\ Phys.\ J.~ A {\bf9}, 35
(2000) and references therein.

\bibitem{Valo00}
A.~Valor, P.-H.~Heenen, and P.~Bonche, Nucl.\ Phys.\ A {\bf671}, 145
(2000) and references therein.

\bibitem{Honn95}
M.~Honma, T.~Mizusaki, and T.~Otsuka, Phys.\ Rev.\ Lett.\ {\bf75},
1284 (1995).

\bibitem{Shim01}
N.~Shimizu, T.~Otsuka, T.~Mizusaki, and M.~Honma, Phys.\ Rev.\ Lett.\
{\bf86}, 1171 (2001).
\end{thebibliography}
\end{document}